\PassOptionsToPackage{hyphens}{url}
\documentclass[aps,physrev,reprint,superscriptaddress,floatfix,raggedbottom]{revtex4-2}

\usepackage{tabularx}
\usepackage{booktabs}
\usepackage{multirow}

\usepackage{amsmath,amssymb,amsfonts}
\usepackage[ruled,lined,linesnumbered]{algorithm2e}
\usepackage{graphicx}
\usepackage{textcomp}
\usepackage{xcolor}
\usepackage{url}

\usepackage{subcaption}
\makeatletter
\long\def\@makecaption#1#2{%
  \vskip\abovecaptionskip
  \begingroup\small
    \leftskip=0pt\relax
    \rightskip=0pt\relax
    \parfillskip=0pt plus 1fil\relax
    \noindent #1: #2\par
  \endgroup
  \vskip\belowcaptionskip}
\makeatother

\usepackage{tikz}
\usepackage{hyperref}

\hypersetup{
  hidelinks,
  pdftitle={Co-Designing Error Mitigation and Error Detection for Logical Qubits},
  pdfauthor={Rohan S. Kumar, Takahiro Tsunoda, Sophia Xue, Dantong Li, Robert J. Schoelkopf, Yongshan Ding},
  pdfsubject={},
  pdfkeywords={}
}

\begin{document}
\raggedbottom

\title{Co-Designing Error Mitigation and Error Detection for Logical Qubits}
\author{Rohan S. Kumar}
\affiliation{Department of Computer Science, Yale University, New Haven, CT, USA}
\affiliation{Yale Quantum Institute, Yale University, New Haven, CT, USA}
\author{Takahiro Tsunoda}
\affiliation{Yale Quantum Institute, Yale University, New Haven, CT, USA}
\affiliation{Department of Applied Physics, Yale University, New Haven, CT, USA}
\author{Sophia Xue}
\affiliation{Yale Quantum Institute, Yale University, New Haven, CT, USA}
\affiliation{Department of Applied Physics, Yale University, New Haven, CT, USA}
\author{Dantong Li}
\affiliation{Department of Computer Science, Yale University, New Haven, CT, USA}
\affiliation{Yale Quantum Institute, Yale University, New Haven, CT, USA}
\author{Robert J. Schoelkopf}
\affiliation{Yale Quantum Institute, Yale University, New Haven, CT, USA}
\affiliation{Department of Applied Physics, Yale University, New Haven, CT, USA}
\author{Yongshan Ding}
\affiliation{Department of Computer Science, Yale University, New Haven, CT, USA}
\affiliation{Yale Quantum Institute, Yale University, New Haven, CT, USA}
\affiliation{Department of Applied Physics, Yale University, New Haven, CT, USA}

\begin{abstract}

Near-term quantum workloads demand error management, yet the two lightest-weight techniques, Quantum Error Detection (QED) and Probabilistic Error Cancellation (PEC), have complementary cost profiles whose joint architectural design space remains unexplored. QED encodes logical qubits and discards error-flagged runs, filtering noise with low qubit overhead but leaving residual errors; PEC can correct these in software, but at exponential cost in noise strength. If QED efficiently reduces per-gate noise, PEC's cost savings can outweigh QED's discard overhead; realizing this, however, requires solving two system-level design challenges.

First, the \textit{QED interval}---how often detection cycles are inserted---is a tunable architectural parameter governing the cost-accuracy tradeoff. We derive an efficiency condition and show that the canonical one-cycle-per-gate frequency does not achieve break-even in any code we evaluate, while optimized intervals on high-rate Iceberg codes do. Second, we discover that naive PEC+QED integration \textit{degrades} accuracy below the QED-only baseline. The root cause is a transient error profile in the first detection cycle that corrupts PEC's noise model. We develop \textit{steady-state extraction}, a co-designed characterization protocol that isolates steady-state error behavior, reducing estimation bias by up to $10.2\times$. On a $[[6,4,2]]$ Iceberg code running QAOA ($p{=}4$--$8$) with a fixed shot budget, PEC+QED achieves $2$--$11\times$ lower absolute error and up to $31\times$ lower MSE versus PEC on physical qubits, with per-interval savings compounding over interval depth.
\end{abstract}

\maketitle

\section{Introduction}\label{sec:intro}

\begin{figure}[t]
  \centering
  \includegraphics[width=0.9\columnwidth]{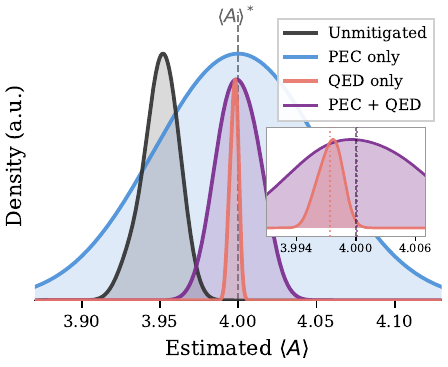}
  \caption{Sampling distribution of the cost estimator
    $\langle A \rangle$ for a MaxCut QAOA
    instance ($p{=}8$) on the $[[6,4,2]]$ Iceberg code under depolarizing noise
    ($p_2 = 10^{-3}$). Curves show CLT-predicted distributions at $N{=}10^3$ shots,
    parameterized by $10^5$-shot empirical statistics. Dashed line: ideal
    $\langle A \rangle^* = 4$. \textbf{Inset:} Zoom near $\langle A \rangle^*$ shows
    that PEC+QED further reduces the residual bias from QED alone, from
    $2.6 \times 10^{-3}$ (QED only) to $9.0 \times 10^{-5}$ (PEC+QED), a
    $29\times$ reduction. Full evaluation in Section~\ref{sec:e2e_eval}.}
  \label{fig:intro-dist}
\end{figure}

\begin{figure*}[t]
  \centering
  \begin{subfigure}[t]{0.48\textwidth}
    \centering
    \includegraphics[width=\linewidth]{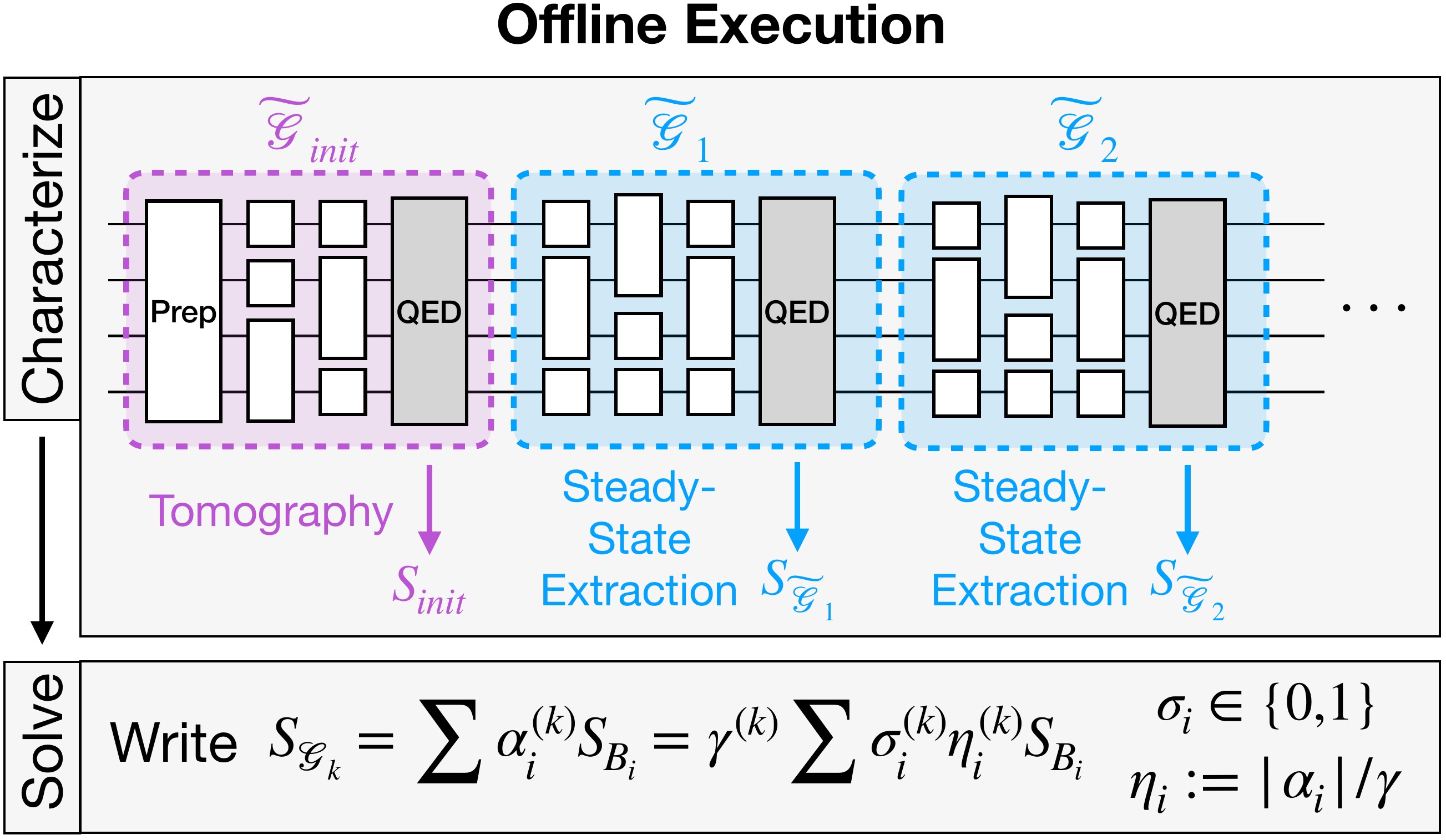}
    \caption{}
    \label{fig:intro-exec-offline}
  \end{subfigure}
  \hfill
  \begin{subfigure}[t]{0.48\textwidth}
    \centering
    \includegraphics[width=\linewidth]{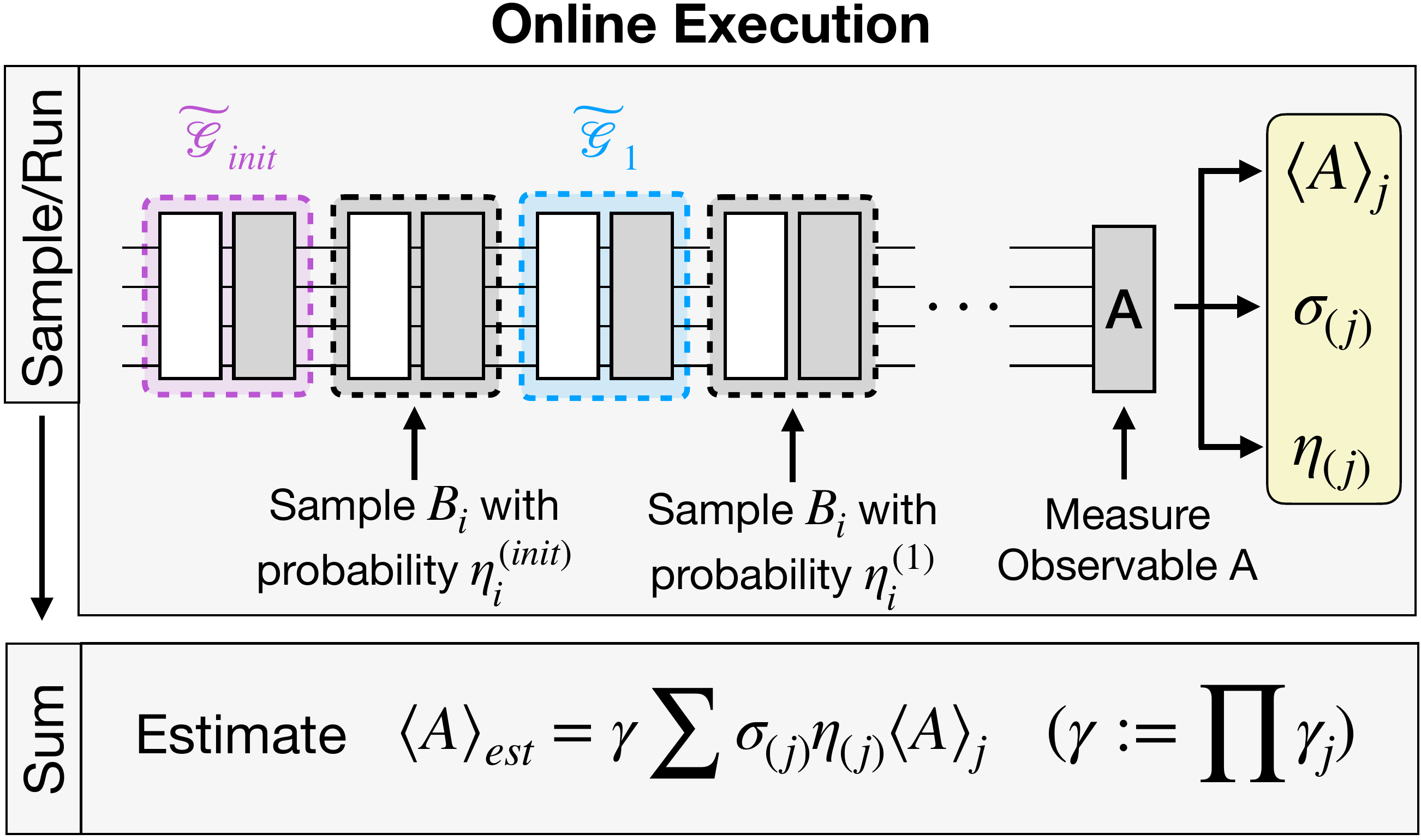}
    \caption{}
    \label{fig:intro-exec-online}
  \end{subfigure}
  \caption{\textbf{PEC+QED co-design architecture.}
  Unlike standard PEC, which mitigates noise gate by gate, this architecture
  operates on \emph{mitigable units}---groups of logical gates bounded by QED
  cycles.
  \textbf{(a)}~Offline: each mitigable unit is characterized offline. The first unit (purple) captures
  the initialization transient; subsequent units (blue) are isolated from
  this transient via \textit{steady-state extraction}
  (Section~\ref{sec:sse}). Each unit's channel is decomposed into a
  quasi-probability distribution over implementable basis operations.
  \textbf{(b)}~Online: PEC stochastically substitutes basis operations
  within each unit according to offline-learned probabilities, combining
  per-shot outcomes into the unbiased estimator
  $\langle A \rangle_{\mathrm{est}}$ with normalization
  $\gamma = \prod_j \gamma_j$.}
  \label{fig:intro-exec}
\end{figure*}

Near-term quantum algorithms for chemistry, optimization, and machine learning require executing circuits with hundreds to thousands of gates on noisy hardware \cite{preskill2018quantum, bharti2022noisy, cerezo2021variational}. Without error management, noise accumulates exponentially with depth, overwhelming any quantum advantage. Quantum Error Correction (QEC) will ultimately enable fault-tolerant computation, but today's devices fall far short of its space and feedback latency requirements. Bridging this gap demands system architectures that co-design lighter-weight error management techniques to extract more from existing hardware. Two promising building blocks are Quantum Error Detection (QED), which encodes logical qubits and discards error-flagged runs \cite{aliferis2007accuracy, gottesman2016quantum}, and Probabilistic Error Cancellation (PEC), which constructs unbiased estimators by stochastically sampling quasi-probability distributions of noisy gates~\cite{temme2017error,endo2018practical}. Both have already demonstrated practical value on current hardware \cite{he2025performance, reichardt2024fault, rines2025demonstration, linke2017fault, vuillot2017error, van2023probabilistic}.

The potential for co-designing these two techniques into a unified system rests on a cost asymmetry. QED leaves residual bias from undetectable errors but incurs only moderate discard overhead. PEC can cancel this residual bias, but its sampling cost grows rapidly with the noise strength it must correct. If QED efficiently reduces the per-interval noise that PEC sees, the savings on PEC's overhead can outweigh QED's discard cost. This work develops an end-to-end system architecture for combining PEC with QED, addressing two key design challenges along the way. As overviewed in Figure~\ref{fig:intro-exec}, the architecture operates in two phases: offline~(\ref{fig:intro-exec-offline}), each mitigable unit (a group of logical gates bounded by QED cycles) is characterized and decomposed into a quasi-probability distribution over basis operations; online~(\ref{fig:intro-exec-online}), PEC samples these distributions to correct each unit, rather than mitigating individual gates. Figure~\ref{fig:intro-dist} previews the practical impact on a QAOA instance: without error management, the observable estimate is sharply biased; PEC removes this bias but amplifies variance; QED has low variance but retains a residual bias from undetectable errors; their co-design enables lower-than-QED bias with lower-than-PEC cost.

\textbf{(1) Sampling advantage via PEC-QED co-design} (Section~\ref{sec:sampling_adv}). To our knowledge, prior PEC+QED work has not quantified the conditions under which PEC on QED codes yields a net sampling advantage, nor identified detection frequency as the architectural parameter governing this tradeoff. In this work, we show that shot savings depend on careful choice of the code parameters, error rates, and how frequently QED cycles are inserted. Concretely, we derive an efficiency condition and show that treating the detection interval as a tunable architectural parameter is essential: the canonical minimum interval ($L{=}1$) does not achieve advantage in any code we evaluate, but optimized intervals enable high-rate Iceberg codes to achieve sampling advantage. Low-rate codes such as the $[[9,1,3]]$ surface code do not reach break-even.

\textbf{(2) Co-designed logical error characterization} (Sections~\ref{sec:nonmarkov}--\ref{sec:sse}). Standard PEC assumes position-independent noise channels, meaning each gate's effective error channel is the same regardless of circuit position, but we quantify how error-detected logical qubits violate this assumption. The first QED cycle exhibits a transient error profile, driven by codespace-leakage injection without prior error removal, that is distinct from the steady-state channel governing subsequent gates (Figure~\ref{fig:intro-exec}). If unaccounted for, this position-dependent error behavior causes PEC to \textit{degrade} fidelity below QED-only baselines. Using the structure of this position dependence, we develop \textit{steady-state extraction}, a characterization protocol co-designed with QED that isolates the steady-state channel by inverting out the first-cycle transient. The protocol applies to any characterization method that yields composable superoperator estimates. In our conservative simulations, this technique reduces observable bias by up to $10.2\times$ below QED-only baselines across depolarizing, amplitude damping, and dephasing noise.

Finally, we validate the complete architecture on a $[[6,4,2]]$ Iceberg code running QAOA with a fixed shot budget at depths $p{=}4$ through $p{=}8$ (Section~\ref{sec:e2e_eval}), demonstrating that PEC+QED achieves $2$--$11\times$ lower absolute error than PEC on physical qubits and up to $31\times$ lower MSE, with the advantage growing as larger QED intervals compound the per-interval $\gamma$ reduction.

At the systems level, the paper contributes a co-designed execution and characterization pipeline that exposes QED interval placement and steady-state-aware channel learning as architecture-level knobs.

\section{Background}\label{sec:background}

\subsection{\label{sec:back_lqc} Logical Quantum Computation}

Quantum computers face fundamental challenges from environmental interactions, control imperfections, and decoherence. A central approach to managing these errors is logical quantum computation, where quantum information is encoded into logical qubits: collections of physical qubits that together form a smaller set of computational qubits protected through redundancy. A logical quantum code is denoted $[[n,k,d]]$, where $n$ physical qubits encode $k$ logical qubits with code distance $d$, meaning the code can detect up to $d-1$ errors.

Two principal paradigms exist for realizing logical qubits: quantum error correction (QEC) and quantum error detection (QED). QEC actively identifies and corrects errors through syndrome measurements and recovery operations, enabling scalable fault-tolerant computation with polynomial overhead when physical error rates fall below a threshold. In contrast, QED identifies errors through syndrome measurements but employs post-selection, discarding computational runs where errors are detected. QED achieves superior error suppression compared to equivalent-distance error-correcting codes at the expense of a shot overhead.

Most codes on digital quantum computers operate within the stabilizer formalism~\cite{gottesman1997stabilizer}, which describes codes through sets of commuting Pauli operators that define a codespace as their mutual +1 eigenspace. Errors manifest as violations of stabilizer conditions. The pattern of stabilizer measurement outcomes, called the syndrome, reveals information about errors without disturbing the encoded quantum information.

\subsection{\label{sec:back_qed} Quantum Error Detection}

Quantum error detection realizes logical qubits by measuring stabilizer syndromes and discarding shots where errors are flagged. Consider a quantum device with gate error rate $\epsilon$ implementing an $[[n, k, d]]$ quantum code with perfect syndrome extraction. The post-selected error rate scales as $\Theta(\epsilon^d)$, comparing favorably to the $\Theta(\epsilon^{\lfloor(d+1)/2\rfloor})$ scaling for QEC codes of equivalent distance. However, each gate incurs a failure probability of $O(\epsilon)$, so for $M$ gates, only a fraction $(1-\epsilon)^M \approx e^{-\epsilon M}$ of shots (single, independent executions of the circuit) successfully pass post-selection.

Beyond favorable error scaling, QED offers practical advantages over QEC. QEC requires passing syndrome information through a decoder to infer and correct errors, necessitating a feedback loop between quantum and classical hardware that is challenging to physically realize. QED circumvents this entirely by avoiding error correction. Additionally, QED enables distance-2 codes capable of detecting single errors without identifying which error occurred. These are codes that cannot function for QEC, which requires distance 3 to correct even a single error. Distance-2 QED codes provide lower physical qubit overhead and high encoding rates, particularly valuable for resource-limited devices.

The $[[k+2, k, 2]]$ Iceberg code \cite{self2024protecting, jin2025iceberg} exemplifies this class of efficient QED codes. With stabilizers $\{X^{\otimes k+2}, Z^{\otimes k+2}\}$, it encodes $k$ logical qubits using only two additional physical qubits. The smallest instance, the $[[4,2,2]]$ code, is the minimal code capable of detecting arbitrary single-qubit errors and has been demonstrated in several break-even experiments \cite{he2025performance, reichardt2024fault, rines2025demonstration, linke2017fault, vuillot2017error}.

\subsection{\label{sec:back_pec} Probabilistic Error Cancellation}

Probabilistic error cancellation (PEC) \cite{temme2017error, endo2018practical, van2023probabilistic, CaiRMP2023} is a quantum error mitigation technique for estimating observable expectation values $\langle \psi | U^\dagger A U | \psi \rangle$ in the presence of noise. This includes many important near-term quantum algorithms: variational quantum eigensolvers (VQE) for molecular ground states \cite{peruzzo2014vqe, mcclean2016theory, kandala2017hardware}, quantum approximate optimization algorithms (QAOA) for combinatorial problems \cite{farhi2014qaoa, farhi2014bounded}, and quantum machine learning applications \cite{mitarai2018qcl, farhi2018qnn, havlicek2019feature}. Noise in the circuit $U$ biases the expectation value away from its ideal value. The objective of PEC is to provide an unbiased estimator at the cost of increased sampling overhead.

PEC operates by expressing the superoperator $S_{\mathcal{G}}$ for an ideal gate operation $\mathcal{G}$ as a linear combination of its implementable noisy counterpart and a set of basis operations: $S_\mathcal{G} = \alpha_0 S_{\tilde{\mathcal{G}}} + \sum_{i=1} \alpha_i S_{\mathcal{B}_i}$, where $S_{\tilde{\mathcal{G}}}$ is the superoperator for the noisy implementation of $\mathcal{G}$, $\{S_{\mathcal{B}_i}\}$ form a complete basis of implementable operations, and the coefficients $\{\alpha_i\}$ constitute a quasi-probability distribution (summing to one but possibly taking negative values). By stochastically sampling operations according to $|\alpha_i|$ and reweighting measurement outcomes by $\mathrm{sgn}(\alpha_i)$, PEC constructs an unbiased estimator of the ideal expectation value. This approach therefore requires full characterization of both the basis set $\{\mathcal{B}_i\}$ and all noisy gate implementations $\tilde{\mathcal{G}}$ in the target circuit. PEC assumes position-independent circuit noise, which permits one-time characterization of each gate type for repeated use in mitigation throughout the circuit.

For bounded observables, the sampling overhead for a circuit scales as $N = O\!\left(\gamma^2 / \delta^2\right)$, where $\gamma = \prod_{g=1}^M \gamma_g$ is the product of quasi-probability norms over all $M$ gates, with each $\gamma_g = \sum_i |\alpha_i^{(g)}|$, and $\delta$ the desired precision; more generally, $\gamma$ captures the dominant exponential overhead while exact finite-shot variance also depends on observable second moments. This quadratic dependence on $\gamma$ makes PEC highly sensitive to noise strength; stronger noise yields larger $|\alpha_i|$ values, increasing the sampling cost. For small physical error rates $\epsilon$, the per-gate overhead is approximated as $\gamma_g \approx (1 + 2\epsilon)$, giving $N = O\!\left((1 + 2\epsilon)^{2M}\right) = O\!\left(e^{4\epsilon M}\right)$. Thus, the overall sampling overhead grows exponentially with the total noise $\epsilon M$, but remains tractable when $\epsilon M = O(1)$.

\subsection{\label{sec:back_characterization} Error Channels and Characterization}
Quantum channels, described mathematically as completely positive trace-preserving (CPTP) maps, characterize how quantum states evolve under noise and operations. Common noise models include depolarizing noise (which randomly applies Pauli errors), amplitude damping (energy dissipation, e.g., qubit decay from $|1\rangle$ to $|0\rangle$), and dephasing (loss of phase coherence without energy loss); this work evaluates all three. These channels can be represented as matrix superoperators that fully specify the transformation. PEC requires accurate channel characterization to construct quasi-probability distributions that precisely cancel the characterized noise.

Several quantum characterization techniques trade completeness against overhead. Cycle benchmarking efficiently extracts error channel properties but does not directly provide complete superoperator descriptions. Gate set tomography (GST) provides thorough characterization with built-in robustness to state preparation and measurement (SPAM) errors, but at substantial experimental cost~\cite{NielsenGST2021, merkel2013self}. Standard quantum process tomography (QPT), while not robust to SPAM errors, offers complete characterization of arbitrary noise channels at lower cost than GST by preparing a tomographically complete set of inputs and performing state tomography on the outputs~\cite{chuang1997prescription}. More recently, circuit-propagation methods leverage structural knowledge of the circuit~\cite{zhong2025combining}: given known or independently characterized physical error rates, these methods reconstruct the total logical noise channel by propagating physical errors through the Clifford circuit structure, avoiding the exponential overhead of logical-level QPT.

\section{Sampling Advantage via PEC+QED Co-design}\label{sec:sampling_adv}

Before combining PEC and QED, we must first determine the conditions under which doing so is worthwhile. The high-level argument is that PEC's shot overhead scales as $e^{4\epsilon M}$ for $M$ gates at per-gate noise $\epsilon$, while QED's discard cost scales as the comparatively smaller $e^{\epsilon M}$. If QED reduces the noise PEC must correct, the exponential savings can dwarf the discard cost, but only if the code is sufficiently efficient at reducing error per shot it discards. As we will show in Section~\ref{sec:interval_eval}, this efficiency depends on how often QED cycles are inserted into the circuit, a tunable parameter we call the \textit{QED interval}. We make this tradeoff precise below by deriving a condition under which PEC on QED codes requires fewer total shots than PEC on the physical system (Section~\ref{sec:adv_condition}), then evaluate it across codes and QED intervals via Stim simulation (Section~\ref{sec:interval_eval}).

\subsection{A Condition for Sampling Advantage}
\label{sec:adv_condition}

Consider an $[[n,k,d]]$ QED code where a QED cycle is applied after each sequence of logical gates. For each such sequence, let the post-selected logical error rate be $\epsilon_\ell$, with rejection probability $p_{\text{rej}}$. Let the equivalent sequence of gates run on physical hardware have error rate $\epsilon_p$. For break-even QED, $\epsilon_\ell < \epsilon_p$. Let there be $M$ such sequences in the circuit. To leading exponential order, the physical-PEC and PEC+QED shot costs are $C_{\mathrm{PEC}}^{\mathrm{phys}} \approx D_p(\mathcal{O})\,e^{4\epsilon_p M}$ and $C_{\mathrm{PEC+QED}}$ $\approx D_\ell(\mathcal{O})\,e^{p_{\text{rej}} M}\,e^{4\epsilon_\ell M}$, where $D_p(\mathcal{O})$ and $D_\ell(\mathcal{O})$ absorb finite-shot prefactors that depend on the observable. Comparing only the dominant exponential terms, the \textit{advantage ratio} is $R = e^{(4\epsilon_p-4\epsilon_\ell-p_{\text{rej}})M}$. Requiring $R > 1$ gives the condition
\begin{equation}
    \frac{\epsilon_p-\epsilon_\ell}{p_{\text{rej}}} > \frac{1}{4}
    \label{eq:efficiency_condition}
\end{equation}

This condition quantifies \textit{QED efficiency}: the error reduction $\epsilon_p - \epsilon_\ell$ achieved per unit of rejection probability.

Under code-capacity assumptions, where logical gates are noisy but syndrome extraction is perfect, the condition is easily satisfied. Let $K$ be the constant number of physical gates required to implement a logical gate, so the pre-detection error rate is $\approx K\epsilon_p$. Since QED cycles are perfect, $\epsilon_\ell = (K\epsilon_p)^d$ and $p_{\text{rej}} = K\epsilon_p$. Substituting into Equation~\ref{eq:efficiency_condition} gives $\epsilon_p < \left(\frac{1 - K/4}{K^d}\right)^{\frac{1}{d-1}}$. For an $[[k+2, k, 2]]$ Iceberg code with $K=1$ and $d=2$, this yields $\epsilon_p < 0.75$; for a $[[9, 1, 3]]$ surface code with $K \approx 3$, it gives $\epsilon_p < 0.10$. These thresholds are far above the break-even threshold for all QED codes.

However, circuit-level noise on QED cycles introduces two effects that degrade the condition:

\textit{1. False Negatives.} Errors that occur within the QED circuit can escape detection, causing increased post-selected error.

\textit{2. False Positives.} These are discarded shots that actually contained no error, which can occur due to both errors on the ancilla qubit during syndrome extraction and measurement errors during post-selection, as described in Appendix~\ref{app:meas_error}.

These effects degrade $\epsilon_\ell$ and $p_{\text{rej}}$ from the code-capacity model. To determine whether realistic codes achieve advantage in practice, and to identify the optimal operating point, we evaluate the efficiency condition across a range of QED intervals using stabilizer simulation.

\subsection{\label{sec:interval_eval} Evaluation: QED Interval Optimization}

\begin{figure*}[!t]
  \centering
  \includegraphics[width=\textwidth]{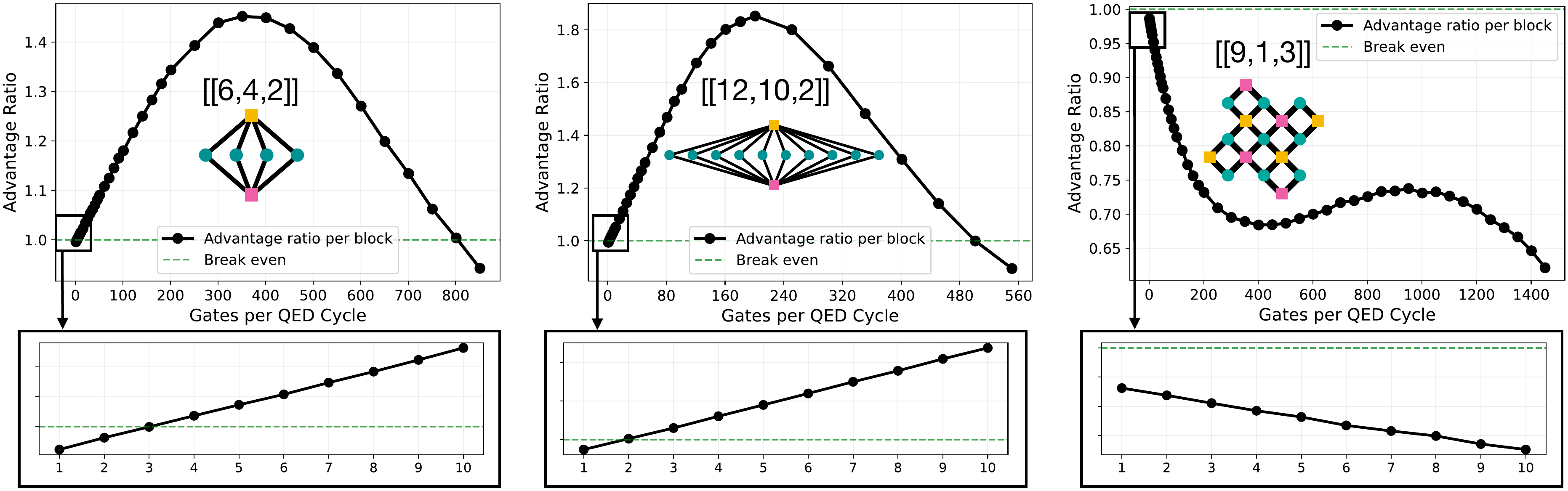}
  \caption{Sampling advantage ratio for the $[[6,4,2]]$, $[[12,10,2]]$, and
    $[[9,1,3]]$ codes. The dominant exponential component of the per-interval
    advantage ratio is plotted as a function
    of the QED interval $L$ (logical gate layers per cycle).
    Results assume depolarizing noise with two-qubit gate error rate $10^{-3}$.
    Values above 1.0 indicate regimes where the dominant cost scaling favors
    PEC+QED over physical-qubit-only PEC.}
  \label{fig:sampling_advantage_results}
\end{figure*}

The condition above applies to any sequence of gates between QED cycles, but the optimal sequence length is far from obvious. Each QED cycle is itself a noisy circuit: while it detects prior errors, it also injects new ones. Too-short sequences pay this overhead repeatedly for little additional detection benefit; too-long sequences let gate errors overwhelm the code. We parameterize the sequence length by the \textit{QED interval} $L$: the number of logical layers executed between consecutive syndrome extraction cycles. The group of $L$ layers bounded by QED cycles forms a \textit{mitigable unit}---the atomic subcircuit that PEC characterizes and mitigates as a single composite operation, replacing per-gate mitigation. Varying $L$ traverses a tradeoff: larger intervals accumulate more gate errors but dilute the fixed overhead of each QED cycle. We now evaluate which codes and intervals achieve sampling advantage in practice, identifying optimal operating points through stabilizer simulation.

\textbf{Implementation Details.}
We evaluate the efficiency metric from Equation~\eqref{eq:efficiency_condition} across QED
intervals using Stim~\cite{gidney2021stim} stabilizer simulations. For a code with QED interval $L$, we construct circuits that:
(1) prepare the logical ground state via noiseless encoding; (2) apply $L$ rounds of depolarizing noise exposure,
where each round corresponds to one logical time step normalized to a two-qubit-gate-time unit;
(3) perform noisy syndrome extraction with post-selection; (4) measure and apply noiseless decoding by reconstructing logical observables from physical measurements (this is done in experiment; e.g. \cite{self2024protecting}). Shots flagging errors during syndrome extraction are discarded. We compute logical error among accepted shots. Our simulations conservatively assign each logical layer one full two-qubit-gate-time unit of noise exposure, regardless of whether the layer contains active operations or idles. If idle layers experience reduced noise in practice (e.g., via dynamical decoupling), the true optimal intervals would be larger and the achievable advantage at least as great. Because the noise model assigns one two-qubit gate per layer, gates and layers are interchangeable throughout this evaluation. Lastly, the first post-initialization QED cycle has a distinct transient error profile (explained in Sections~\ref{sec:nonmarkov}--\ref{sec:sse}). We account for this by estimating the steady-state per-interval logical error and rejection rates by removing the initialization-cycle contribution via a reference measurement.

\textbf{Results.}
Figure~\ref{fig:sampling_advantage_results} shows the per-interval advantage ratio $e^{(4\epsilon_p-4\epsilon_\ell-p_{\text{rej}})}$ as a function of the QED interval $L$.
We evaluate three representative codes: the $[[6,4,2]]$ and $[[12,10,2]]$ Iceberg codes, and the $[[9,1,3]]$ surface code, for depolarizing noise with two-qubit gate error rate $p=10^{-3}$. Throughout all experiments in this work, the single-qubit error rate is one-tenth the two-qubit error rate, consistent with experimental demonstrations \cite{google2025quantum, bluvstein2024logical, ransford2025helios}. We select the $[[6,4,2]]$ and $[[12,10,2]]$ codes because their logical qubit counts span the practical range for characterization: QPT scales to $\sim$4 logical qubits and cycle benchmarking to $\sim$10.

The results reveal three key insights. First, the minimum interval ($L=1$) does not achieve sampling advantage for any code in our evaluation. Second, high-rate codes exhibit fundamentally different efficiency profiles than low-rate codes. The $[[12,10,2]]$ Iceberg code, which encodes 10 logical qubits using only 12 physical qubits, achieves break-even overhead at 2 logical layers per QED cycle, whereas the $[[9,1,3]]$ surface code, which dedicates 9 physical qubits to encode a single logical qubit, degrades in shot efficiency for larger intervals. This is expected: low-rate codes contain many more physical than logical qubits, so $\epsilon_\ell$ grows much more rapidly than $\epsilon_p$. Third, advantageous codes exhibit an optimal QED interval $L_{\text{max}}$ that maximizes the dominant exponential advantage, and a threshold interval $L_{\text{thresh}}$ at which the exponential component first exceeds break-even. The $[[6,4,2]]$ code has $(L_{\text{thresh}}, L_{\text{max}}) = (4, 345)$ layers. The $[[12,10,2]]$ code yields $(L_{\text{thresh}}, L_{\text{max}}) = (2, 208)$ layers. Because $L_{\text{max}}$ exceeds practicality relative to the size of these codes, the dominant exponential term improves monotonically with $L$ across all evaluated intervals. The $[[9,1,3]]$ code does not achieve advantage at any interval, demonstrating that high-rate codes are essential for PEC+QED cost savings. These per-interval factors characterize the dominant exponential component of PEC's sampling cost.

\section{Position-Dependent Logical Error Behavior in Error-Detected Qubits}\label{sec:nonmarkov}

The sampling advantage results of Section~\ref{sec:sampling_adv} establish \textit{when} PEC+QED saves shots, but applying PEC requires accurate characterization of the noise channel. Standard PEC assumes that this channel is position-independent: the same for every gate regardless of position in the circuit. We now show that error-detected logical qubits violate this assumption: the error channel after the first QED cycle differs fundamentally from the steady-state channel governing all subsequent gates. This position-dependent behavior, if unaccounted for, causes PEC to \textit{degrade} performance below QED-only baselines (as we demonstrate in Section~\ref{sec:qutip_eval}). Understanding its origin is therefore essential for making PEC+QED work in practice.

\subsection{\label{sec:eigenstructure} Transition Matrix Model and Eigenstructure}

To understand the origin of this behavior, we model each noisy QED cycle as a stochastic transition matrix on Hilbert-space basis states. The matrix $T_p$ captures how state populations evolve under error rate $p$: Pauli faults are inserted at each error location in the Clifford syndrome extraction circuit, propagated to the circuit output, and averaged over all fault configurations. Experimental post-selection is modeled by aggregating all flagged syndrome states into a single absorbing reject state, yielding a reduced matrix $T_p^{\mathrm{abs}}$ on accepted states plus the absorbing state. We define the \textbf{Accepted State Correctness (ASC)} as the conditional probability of remaining in the correct logical state given survival through post-selection; it proxies quantum fidelity and directly bounds observable bias. We summarize the key results here; Appendix~\ref{app:tm} provides comprehensive details.

Spectral analysis of $T_p^{\mathrm{abs}}$ reveals a clean separation into fast and slow timescales that directly explains the position-dependent behavior. For notational simplicity, we denote this matrix as $T$. For small error probability $p$, standard perturbation theory~\cite{bauer1960norms,seneta1981nonnegative} yields:
\begin{equation}
\label{eq:tm_eigenvals}
    \sigma(T) = \{1\} \cup \{1 - \alpha_i p + O(p^2)\}_{i=1}^{2^k} \cup \{\beta_j p + O(p^2)\}_{j=1}^{2^n - 2^k},
\end{equation}
where $\alpha_i, \beta_j > 0$ are code- and circuit-specific constants. This spectrum defines three groups: one exactly stationary mode (the absorbing state), $2^k$ ``slow'' modes with eigenvalues $1 - O(p)$, and $2^n - 2^k$ ``fast'' modes with eigenvalues $O(p)$.

\textbf{Consequences for ASC evolution.}
Because $T$ is diagonalizable, the state after $r$ rounds evolves as a linear combination of eigenmodes, each weighted by $\lambda_i^r$. Fast modes with eigenvalues $O(p)$ contribute significantly in the first QED cycle but vanish within $O(1)$ rounds. Subsequent cycles are dominated by slow modes with eigenvalues $1 - O(p)$, producing steady-state decay behavior. Since ASC inherits this multi-exponential structure, the first QED cycle operates in a transient regime while all subsequent cycles experience the steady-state channel. The transition matrix simulations in Section~\ref{sec:tm_sim} confirm these spectral predictions numerically.

\subsection{\label{sec:physical_interp} Physical Interpretation}

The ASC provides a phaseless proxy for quantum fidelity that directly bounds observable measurement bias. Our simulations at the Hamiltonian level validate this proxy: both phase-aware fidelity (Figure~\ref{subfig:nonstat_codes}) and phaseless ASC exhibit identical fast/slow decay structure, confirming that the position-dependent ASC dynamics directly determine characterization requirements for mitigating observable errors.

These findings have a simple physical interpretation. Errors within syndrome extraction circuits can produce states outside the logical codespace that nonetheless pass post-selection; these undetected leakage states occur with probability $O(p)$ regardless of code distance due to the existence of at least one error location in the QED circuit where this can occur. While subsequent QED cycles will likely detect these states, each cycle also introduces new $O(p)$ errors. The critical asymmetry occurs in the first cycle: with no prior errors to detect, the initial syndrome extraction only \textit{injects} leakage without removing any, creating an $O(p)$ transient population. Subsequent cycles reach quasi-steady state, where error detection and injection balance at comparable rates. This startup transient fundamentally distinguishes first-cycle dynamics from steady-state operation, similar to cache warmup effects in classical systems. Section~\ref{sec:sse} develops a characterization protocol to account for this effect.

\subsection{\label{sec:tm_sim} Simulating the $[[4,2,2]]$ Transition Matrix}

\begin{table}[t]
\centering
\small
\begin{tabularx}{\columnwidth}{lccccc}
\toprule
\textbf{2Q error} & $\boldsymbol{\lambda_1}$ & $\boldsymbol{\lambda_2}$ & $\boldsymbol{\lambda_3}$ & $\boldsymbol{\lambda_4}$ & \textbf{Fast Lifetime} \\
\midrule
$1.0\times10^{-4}$ & 1.0000 & 0.9976 & 0.9976 & 0.0002 & 0.1174 \\
$5.0\times10^{-4}$ & 1.0000 & 0.9881 & 0.9881 & 0.0010 & 0.1447 \\
$1.0\times10^{-3}$ & 1.0000 & 0.9763 & 0.9763 & 0.0020 & 0.1606 \\
$5.0\times10^{-3}$ & 1.0000 & 0.8882 & 0.8873 & 0.0095 & 0.2147 \\
\bottomrule
\end{tabularx}
\caption{Eigenvalues of the $[[4,2,2]]$ Transition Matrix as a function of 2Q error probability. ``Fast Lifetime'' is the characteristic timescale of the fast-decay eigenvalue $\lambda_4$, which is always less than 1.}
\label{tab:tm_data}
\end{table}

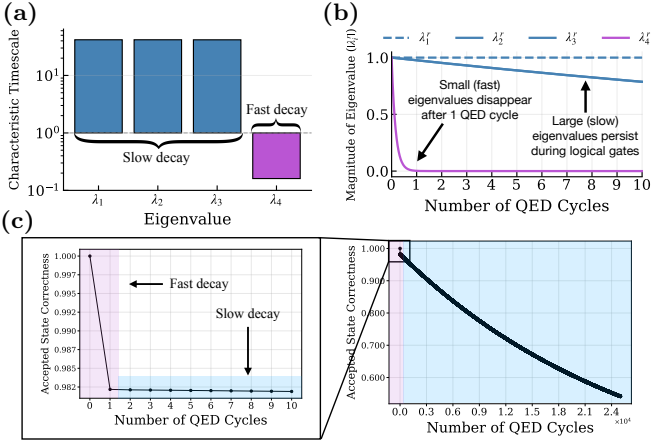
\begin{figure}[!t]
  \centering
  \begin{minipage}[t]{0.48\columnwidth}
    \centering
    \begin{tikzpicture}[baseline]
      \node[anchor=north west,inner sep=0] (img)
        {\includegraphics[width=\linewidth]{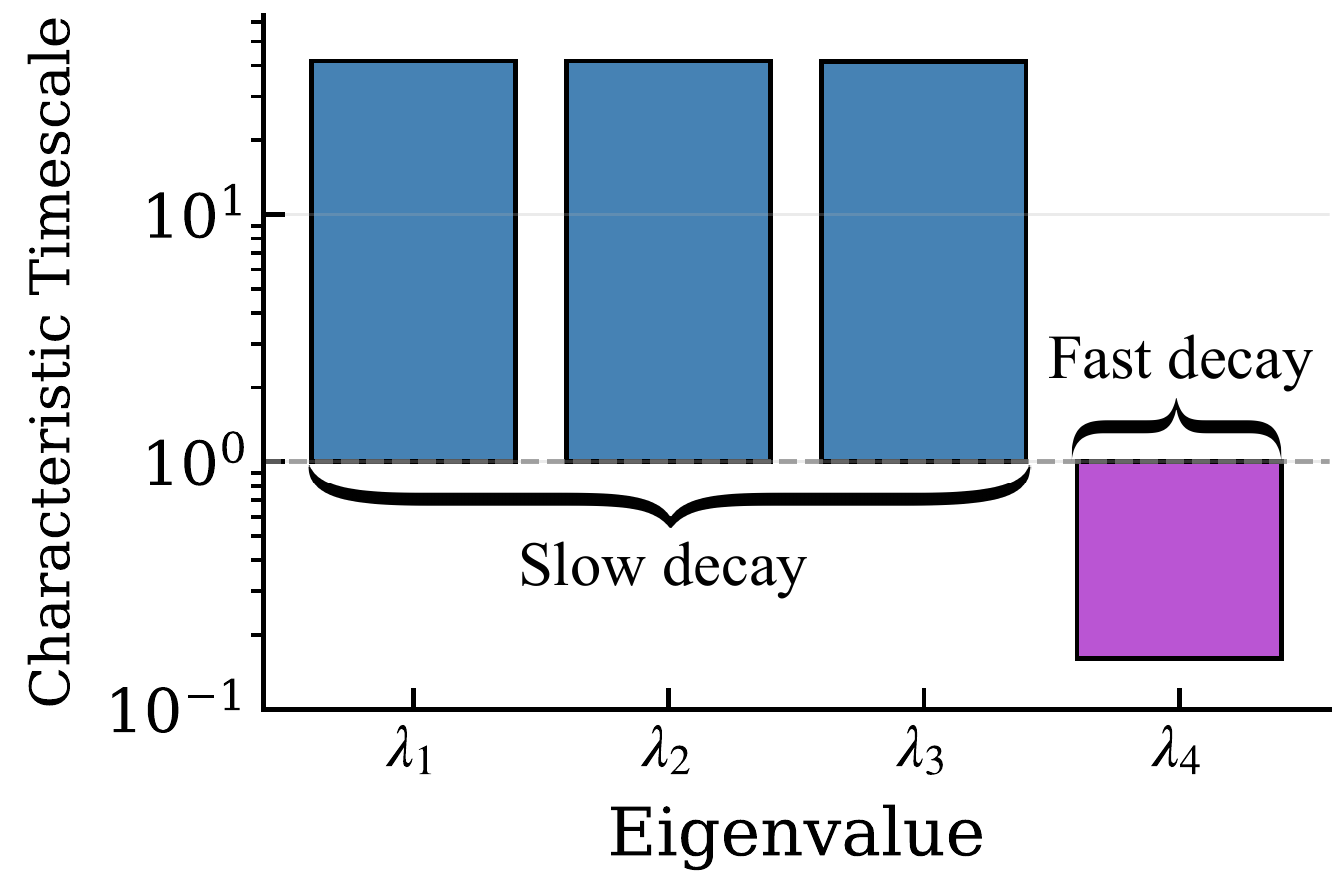}};
      \node[overlay, anchor=north west, xshift=-3pt, yshift=12pt, font=\bfseries]
        at (img.north west) {(a)};
    \end{tikzpicture}
  \end{minipage}%
  \hfill
  \begin{minipage}[t]{0.48\columnwidth}
    \centering
    \begin{tikzpicture}[baseline]
      \node[anchor=north west,inner sep=0] (img)
        {\includegraphics[width=\linewidth]{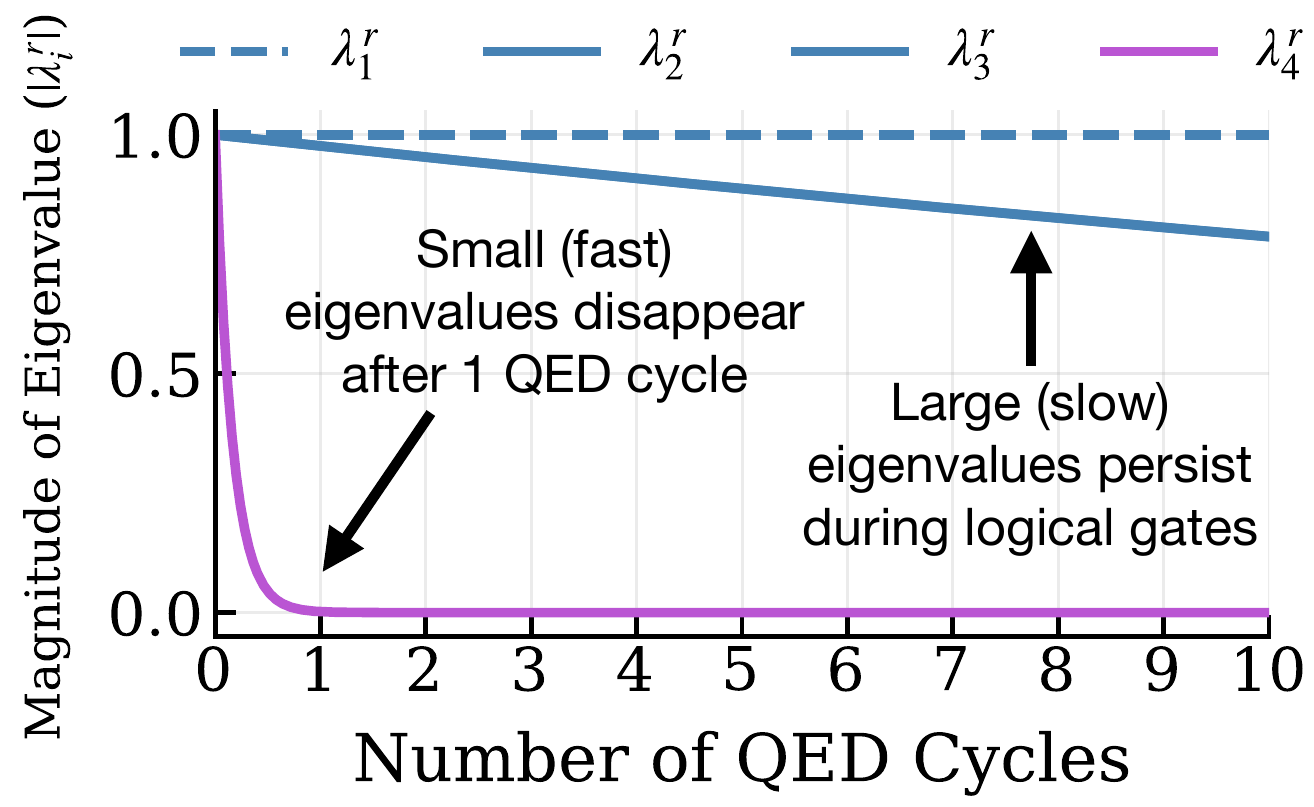}};
      \node[overlay, anchor=north west, xshift=-10pt, yshift=12pt, font=\bfseries]
        at (img.north west) {(b)};
    \end{tikzpicture}
  \end{minipage}

  \begin{minipage}[t]{0.95\columnwidth}
    \centering
    \begin{tikzpicture}[baseline]
      \node[anchor=north west,inner sep=0] (img)
        {\includegraphics[width=\linewidth]{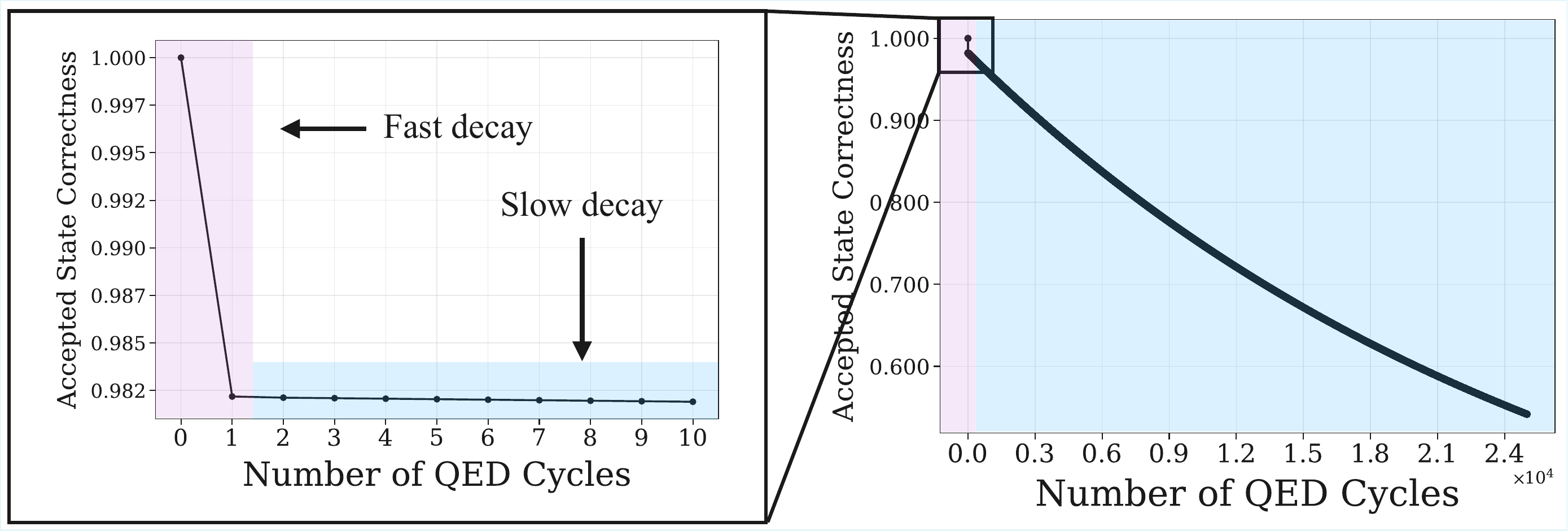}};
      \node[overlay, anchor=north west, xshift=-10pt, yshift=12pt, font=\bfseries]
        at (img.north west) {(c)};
    \end{tikzpicture}
  \end{minipage}

  \caption{Transition Matrix data for the $[[4,2,2]]$ code with depolarizing two-qubit gate error rate $p=10^{-3}$. \textbf{(a)} Characteristic timescales of eigenvalues, given by $-1/\ln(\lambda_i)$, reflect decay time for each mode. The fast-decay $\lambda_4$ eigenvalue is much smaller than the others, corresponding to a fast decay mode. \textbf{(b)} Eigenvalue decay $\lambda_i^r$ after $r$ QED cycles. The fast decay mode becomes negligible after one cycle. \textbf{(c)} Accepted-State Correctness (ASC) over QED cycles. The initial drop is caused by the fast decay mode. Subsequent smooth evolution over $O(10^4)$ timesteps, where logical gates operate, is governed by slow decay modes.}
  \label{fig:tm_ASP}
\end{figure}

We validate these predictions on the $[[4,2,2]]$ code (the smallest QED code) for interpretable transition-matrix construction. We reduce the basis states to four symmetry classes---detected errors, correct logical state, incorrect logical states, and undetected leakage---and compute the eigenvalues of the reduced matrix (see Appendix~\ref{app:tm} for construction details). As shown in Table~\ref{tab:tm_data}, these eigenvalues match the structure found in Equation~\ref{eq:tm_eigenvals} with $\lambda_1 = 1$, $\lambda_2,  \lambda_3 = 1-\Theta(p)$, and $\lambda_4 =\Theta(p)$. For a wide range of realistic $p$, the fast-decay transient always decays in at most 1 step, validating that it can be effectively isolated within a single cycle. Figure~\ref{fig:tm_ASP} displays ASC data for $p=10^{-3}$: the ASC exhibits significant transient dynamics in the first cycle, while subsequent cycles depend almost exclusively on the slow decay modes governing steady-state operation. Section~\ref{sec:sse} develops a characterization protocol inspired directly by these dynamics.

\textbf{Hamiltonian-level validation across codes.} To validate beyond the stabilizer-level transition matrix analysis, we simulate repeated QED cycles using Hamiltonian-level Lindblad master equation simulation via QuTiP~\cite{johansson2012qutip} on the $[[5,1,3]]$, $[[4,2,2]]$, and $[[4,1,2]]$ codes. As shown in Figure~\ref{subfig:nonstat_codes}, all codes exhibit the predicted fast-decay transient in the first cycle followed by consistent slow-decay dynamics in subsequent cycles, confirming that the two-timescale behavior is a generic property of QED codes rather than an artifact of a specific code or noise model.

\section{Steady-State Extraction}\label{sec:sse}

\subsection{\label{sec:naive_fails} Why Naive Characterization Fails}

The position-dependent dynamics from Section~\ref{sec:nonmarkov} create a critical mismatch for standard characterization. When we characterize a logical gate $\widetilde{\mathcal{G}}$ that occurs after the first QED cycle using standard process tomography, we must: (1) initialize the logical state, (2) apply $\widetilde{\mathcal{G}}$, and (3) perform syndrome extraction. The resulting characterized superoperator (call it $S^{\text{naive}}_{\widetilde{\mathcal{G}}}$) describes this entire sequence and unavoidably captures the fast-decay transient. During circuit execution, however, the same gate $\widetilde{\mathcal{G}}$ operates on an initialized system that has passed previous QED cycles and exhibits only steady-state dynamics.

The consequence is severe: regardless of whether logical initialization includes a QED cycle during tomography, naive characterization captures fast-decay dynamics. If initialization includes QED, the fast decay appears there; if not, it appears in the post-gate QED cycle (which is unavoidable). Either way, $S^{\text{naive}}_{\widetilde{\mathcal{G}}}$ includes transient dynamics that do not reflect the gate's actual steady-state behavior during computation. As we demonstrate in Figure~\ref{fig:full_sim_results}(d)--(f), applying PEC with these mischaracterized channels \textit{degrades} fidelity below QED-only baselines, making the mitigation counterproductive.

\subsection{\label{sec:sse_protocol} The Steady-State Extraction Protocol}

\begin{figure}[!t]
  \centering
  \begin{tikzpicture}[baseline]
    \node[anchor=north west,inner sep=0] (img)
      {\includegraphics[width=\columnwidth]{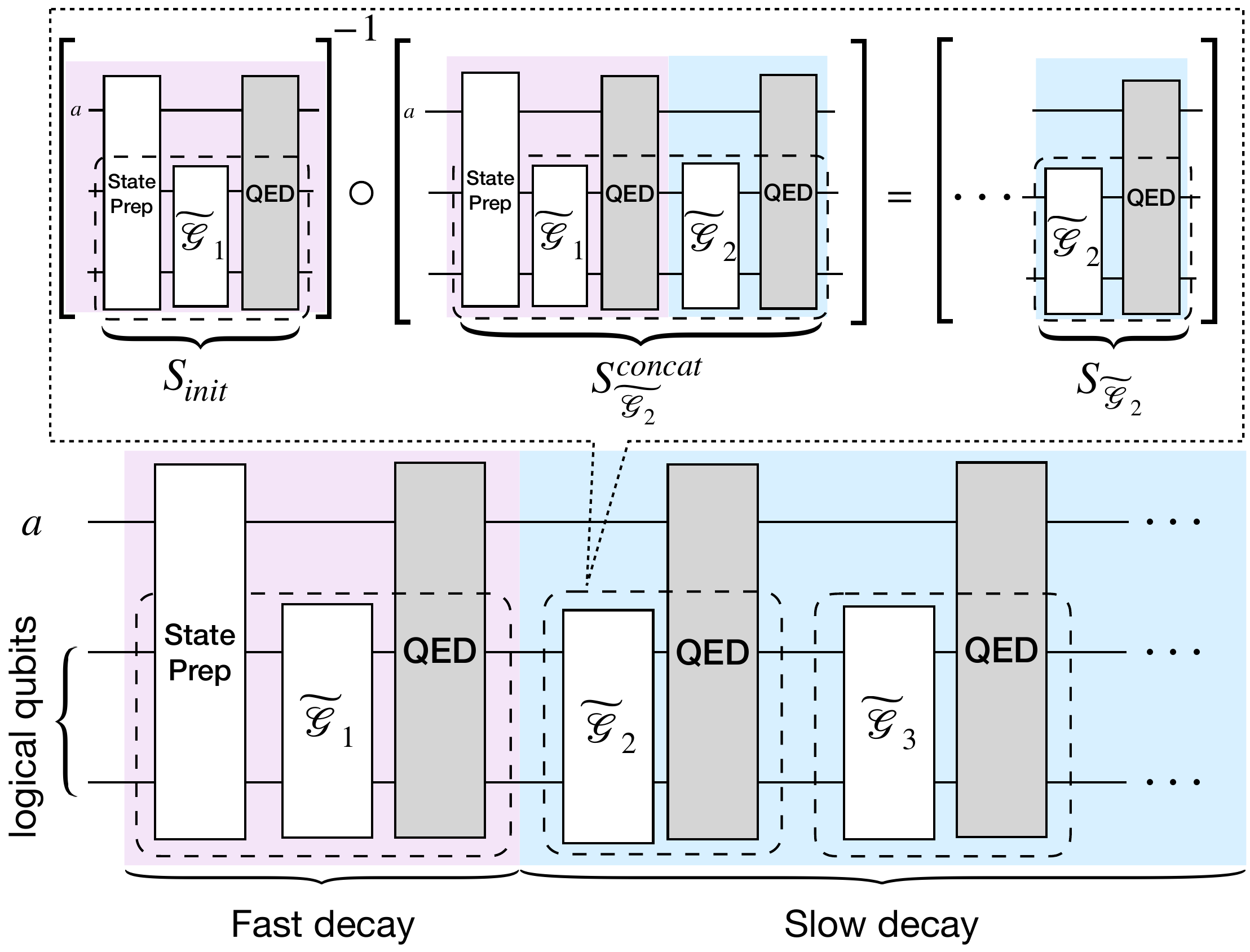}};
  \end{tikzpicture}
\caption{\textbf{Steady-State Extraction Protocol.} Logical circuits exhibit fast-decay dynamics during initialization (purple) before settling into steady-state operation (blue). During characterization, superoperator inversion removes the effect of the initial transient, yielding accurate steady-state gate channels for PEC. State preparation error is naturally absent from all but the first mitigable unit, which spans from initialization through the first QED cycle.}
\label{fig:sse_protocol}
\end{figure}

We resolve this issue through a two-stage process (Figure~\ref{fig:sse_protocol}). The key insight is that during characterization, we can isolate the fast-decay transient into a well-defined initialization subcircuit and invert its effect from gate superoperators. To construct this reference subcircuit, we define the characterization initialization to include a QED cycle following state preparation: if logical initialization uses a unitary encoder circuit, we append a QED cycle; if initialization relies on post-selected syndrome measurements, the required QED cycle is already inherent.
First, we separately characterize $S_{\text{init}}$: the superoperator spanning from logical state preparation through the first QED cycle. \textit{This is the only extra overhead incurred by steady-state extraction}, and causes minimal overhead since it only needs to be characterized once. Second, for each gate $\widetilde{\mathcal{G}}$, we characterize $S_{\text{init}}$ composed with the gate and its associated QED cycle, which we denote $S^{\text{concat}}_{\widetilde{\mathcal{G}}}$.

By the definitions of each superoperator, we can decompose the concatenated sequence as $S^{\text{concat}}_{\widetilde{\mathcal{G}}} = S_{\widetilde{\mathcal{G}}} \circ S_{\text{init}}$, where $S_{\widetilde{\mathcal{G}}}$ is the desired steady-state channel. From this, we can extract $S_{\widetilde{\mathcal{G}}}$ using directly characterizable superoperators~\cite{qin2025classical}:
\begin{equation}
S_{\widetilde{\mathcal{G}}} = S^{\text{concat}}_{\widetilde{\mathcal{G}}} \circ S_{\text{init}}^{-1}.
\label{eq:superop_inversion}
\end{equation}
This process removes the fast-decay transient, yielding an accurate characterization of the slow-decay channel that dominates logical computation. The resulting superoperators correctly describe runtime gate behavior and enable effective PEC mitigation. Moreover, because $S_{\text{init}}$ is independently characterized, the first mitigable unit is itself directly mitigable by PEC, rather than serving only as a reference for characterizing subsequent units.

\textbf{Logical-level and physical-level instantiations.} Steady-state extraction can be applied at two levels of abstraction, each with distinct tradeoffs.

In the \textit{logical-level} approach, one applies QPT directly to logical gates and inverts the initialization superoperator in logical space. This provides a direct route to the noise model without physical-level tomography. However, inverting at the logical level can omit valuable information about the structure of logical leakage, potentially causing lower-quality superoperator reconstructions. We evaluate this approach as an adversarial stress test in Section~\ref{sec:qutip_eval}, and show that it still produces high-quality reconstructions, achieving up to a $10.2\times$ reduction in error when combined with PEC.

In the \textit{physical-level} approach, one characterizes errors at the physical qubit level (for example, using the propagation-based method described in Section~\ref{sec:back_characterization}) and inverts the first-cycle transient in the full physical Hilbert space before projecting to logical space. The physical-level approach is more natural for the position-dependent error problem because the leakage states that drive the transient are clearly resolved as distinct basis states at the physical level, whereas at the logical level their effects are averaged into the logical channel. The cost is a higher-dimensional inversion: one inverts superoperators of dimension $4^n$ in physical space rather than $4^k$ in logical space. For high-rate codes (e.g. Iceberg codes in our work), this overhead is modest (the $[[6,4,2]]$ code adds only 2 physical qubits beyond the 4 logical qubits), and low-rate codes where the overhead would be prohibitive do not achieve sampling advantage in any case (Section~\ref{sec:adv_condition}). We validate the physical-level approach in the end-to-end evaluation of Section~\ref{sec:e2e_eval}.

\textbf{Why this works on QED systems.}
Inverting out prior subcircuits is generally problematic on physical qubits because characterizing any preceding circuit segment conflates all error sources in that segment, including state preparation and measurement (SPAM) errors, which are often significant. This makes clean inversion impossible without expensive protocols like gate set tomography. In contrast, QED codes provide a crucial structural advantage: because post-selection rejects any shot with a non-trivial syndrome, a single measurement fault flips a syndrome bit and triggers rejection rather than corrupting the post-selected logical channel; corrupting accepted shots requires coincident data and measurement errors, a higher-order event (Appendix~\ref{app:meas_error}). This measurement error suppression, validated empirically in Figure~\ref{subfig:qed_meas_supremacy} ($7.67\times$ lower prediction error for logical-level versus physical-level characterization), enables the superoperator inversion to cleanly isolate the fast-decay transient. A key side effect is that the inversion also removes state preparation errors from the characterized gate channels, a component of SPAM that typically requires expensive protocols like gate set tomography (GST) to handle~\cite{Combes2017}. Steady-state extraction achieves this as a natural byproduct: by inverting out $S_{\text{init}}$, both the first-cycle transient and any state preparation imperfections are eliminated from the gate characterization, without the full overhead of GST. At runtime, state preparation error is naturally absent from all but the first mitigable unit, which spans from initialization through the first QED cycle. Because $S_{\text{init}}$ is independently characterized, PEC can also mitigate this first unit directly. The net effect is leading-order SPAM suppression: the inversion removes state preparation error from subsequent gate characterizations, while measurement error suppression ensures that the characterization of $S_{\text{init}}$ cleanly captures state preparation dynamics, enabling PEC to correct them directly in the first unit. Although we instantiate steady-state extraction with QPT and circuit-propagation methods, the principle of isolating first-cycle transients during characterization applies to any characterization protocol that yields composable superoperator estimates.

\subsection{\label{sec:qutip_eval} Evaluation: Hamiltonian-Level Validation}

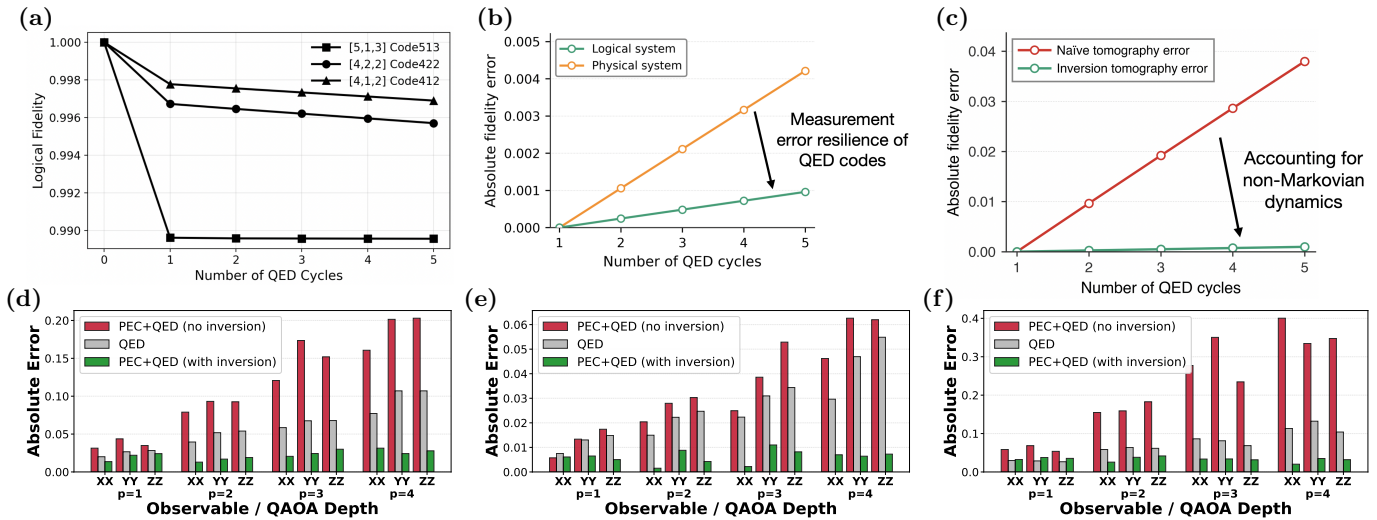
\begin{figure*}[!t]
  \centering
  \begin{minipage}[t]{0.32\textwidth}
    \centering
    \begin{tikzpicture}[baseline]
      \node[anchor=north west,inner sep=0] (img)
        {\includegraphics[width=\linewidth]{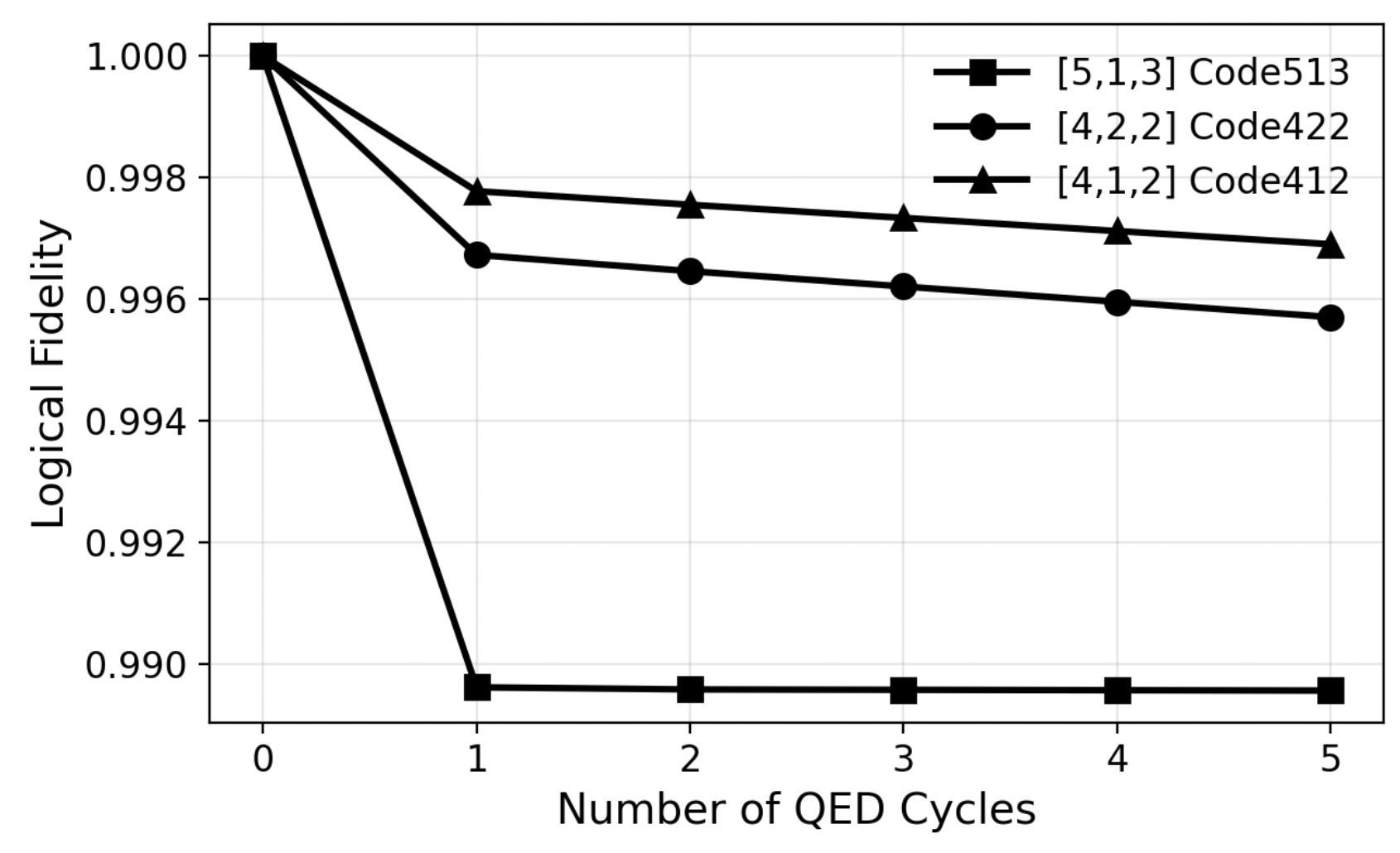}};
      \node[overlay, anchor=north west, xshift=-5pt, yshift=10pt, font=\bfseries]
        at (img.north west) {(a)};
    \end{tikzpicture}
    \phantomsubcaption\label{subfig:nonstat_codes}
  \end{minipage}
  \hfill
  \begin{minipage}[t]{0.32\textwidth}
    \centering
    \begin{tikzpicture}[baseline]
      \node[anchor=north west,inner sep=0] (img)
        {\includegraphics[width=\linewidth]{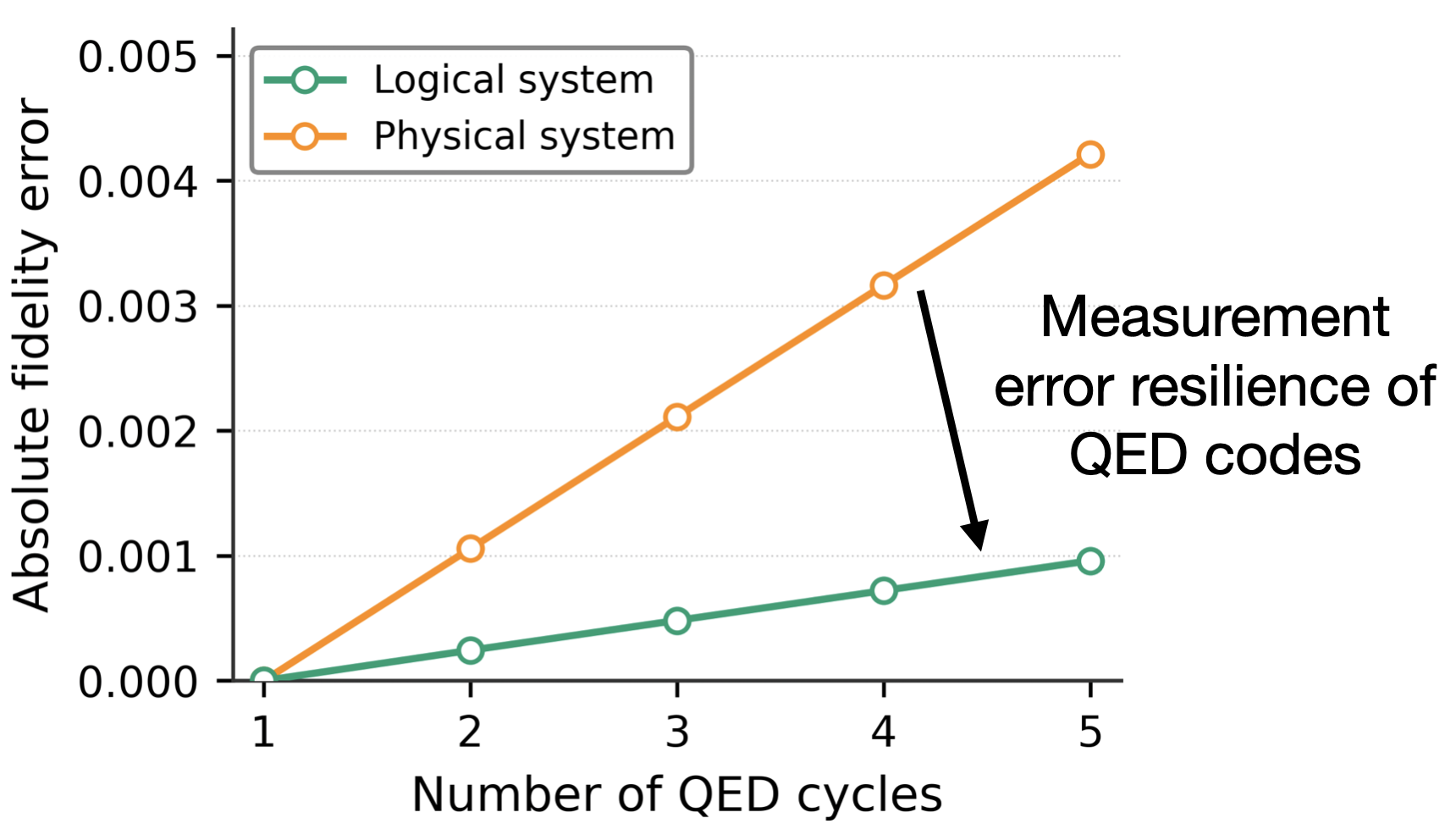}};
      \node[overlay, anchor=north west, xshift=-5pt, yshift=10pt, font=\bfseries]
        at (img.north west) {(b)};
    \end{tikzpicture}
    \phantomsubcaption\label{subfig:qed_meas_supremacy}
  \end{minipage}
  \hfill
  \begin{minipage}[t]{0.32\textwidth}
    \centering
    \begin{tikzpicture}[baseline]
      \node[anchor=north west,inner sep=0] (img)
        {\includegraphics[width=\linewidth]{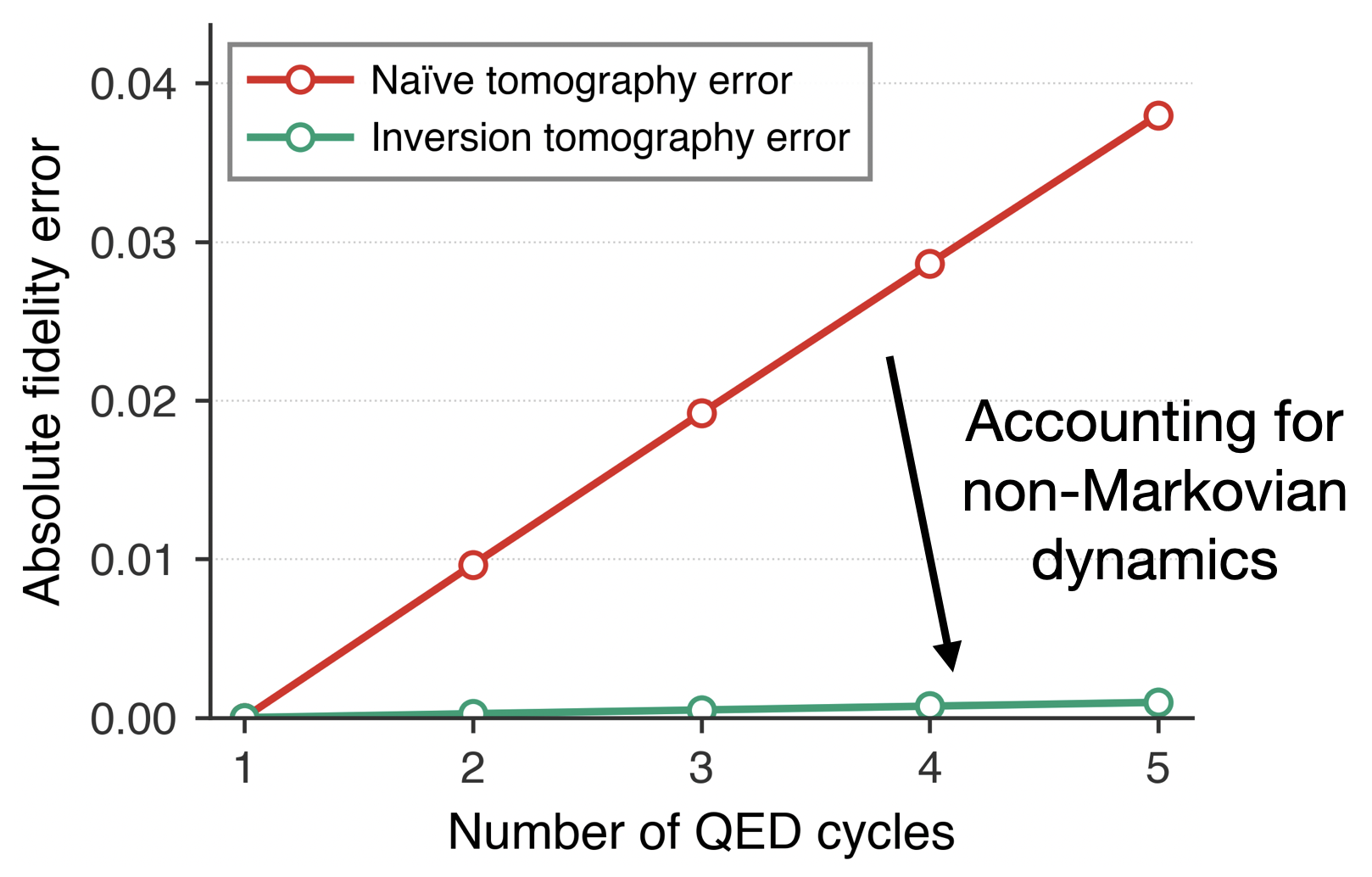}};
      \node[overlay, anchor=north west, xshift=-5pt, yshift=10pt, font=\bfseries]
        at (img.north west) {(c)};
    \end{tikzpicture}
    \phantomsubcaption\label{subfig:inv_supremacy}
  \end{minipage}

  \begin{minipage}[t]{0.32\textwidth}
    \centering
    \begin{tikzpicture}[baseline]
      \node[anchor=north west,inner sep=0] (img)
        {\includegraphics[width=\linewidth]{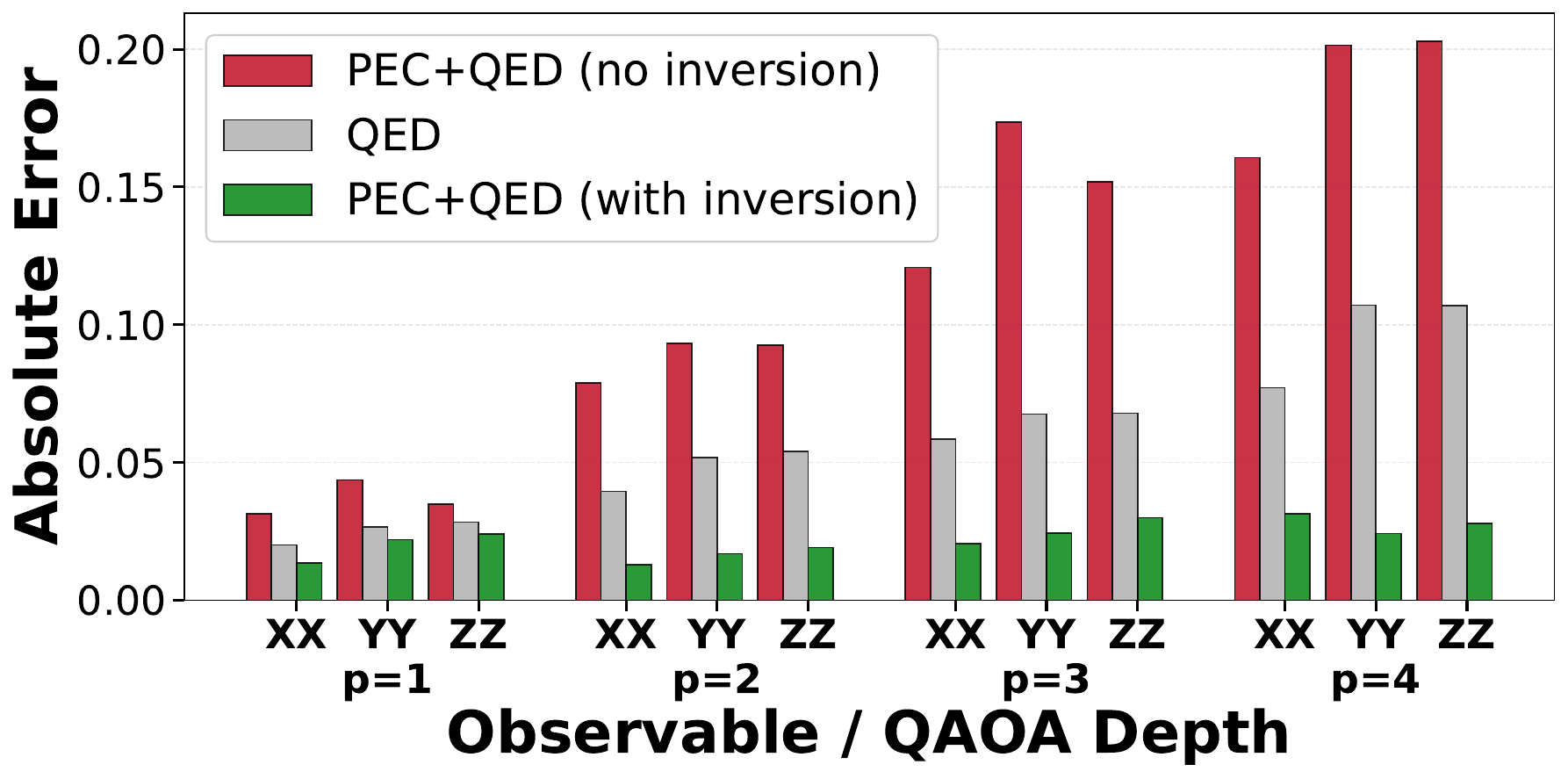}};
      \node[overlay, anchor=north west, xshift=-10pt, yshift=10pt, font=\bfseries]
        at (img.north west) {(d)};
    \end{tikzpicture}
  \end{minipage}
  \hfill
  \begin{minipage}[t]{0.32\textwidth}
    \centering
    \begin{tikzpicture}[baseline]
      \node[anchor=north west,inner sep=0] (img)
        {\includegraphics[width=\linewidth]{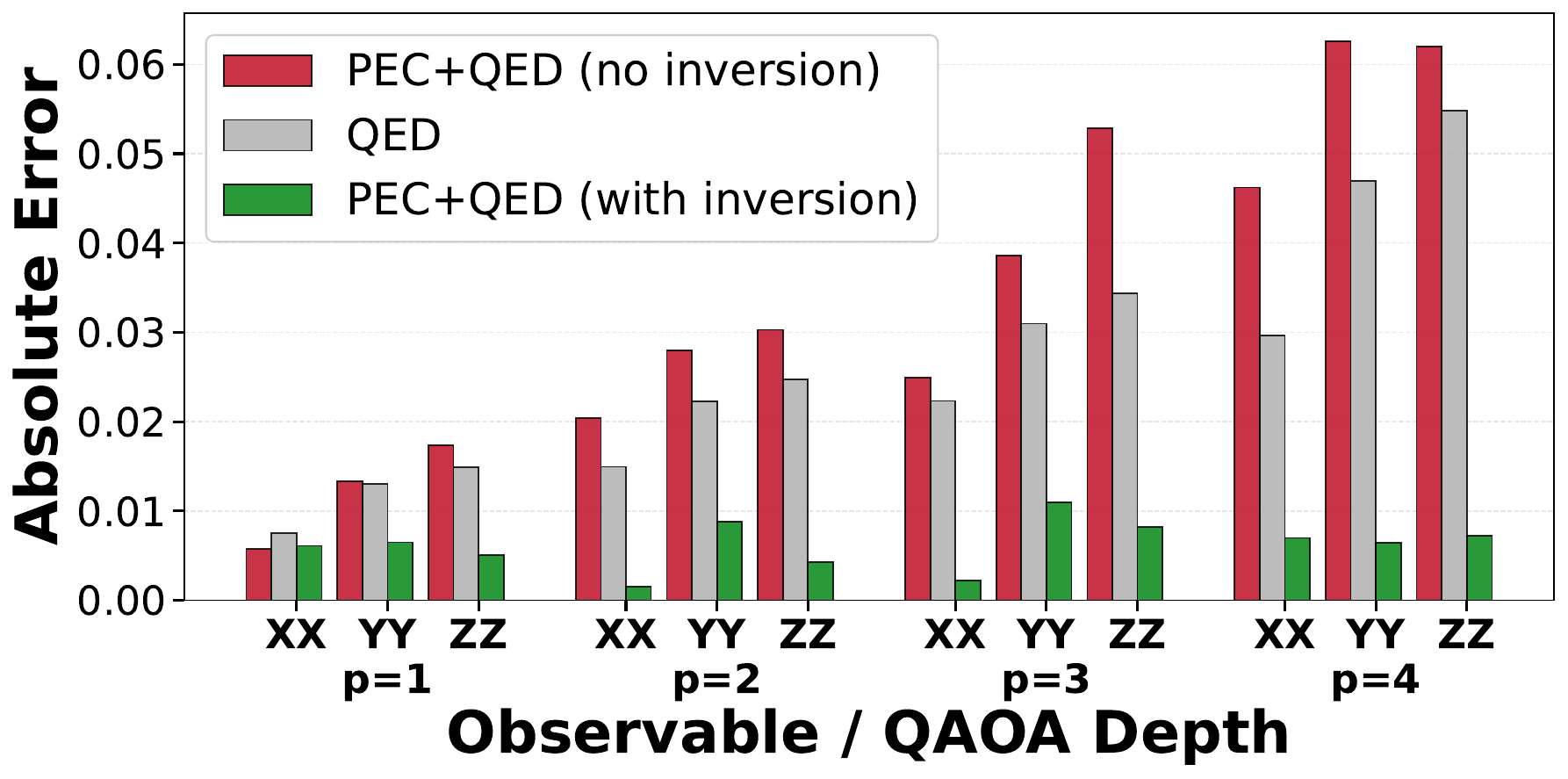}};
      \node[overlay, anchor=north west, xshift=-10pt, yshift=10pt, font=\bfseries]
        at (img.north west) {(e)};
    \end{tikzpicture}
  \end{minipage}
  \hfill
  \begin{minipage}[t]{0.32\textwidth}
    \centering
    \begin{tikzpicture}[baseline]
      \node[anchor=north west,inner sep=0] (img)
        {\includegraphics[width=\linewidth]{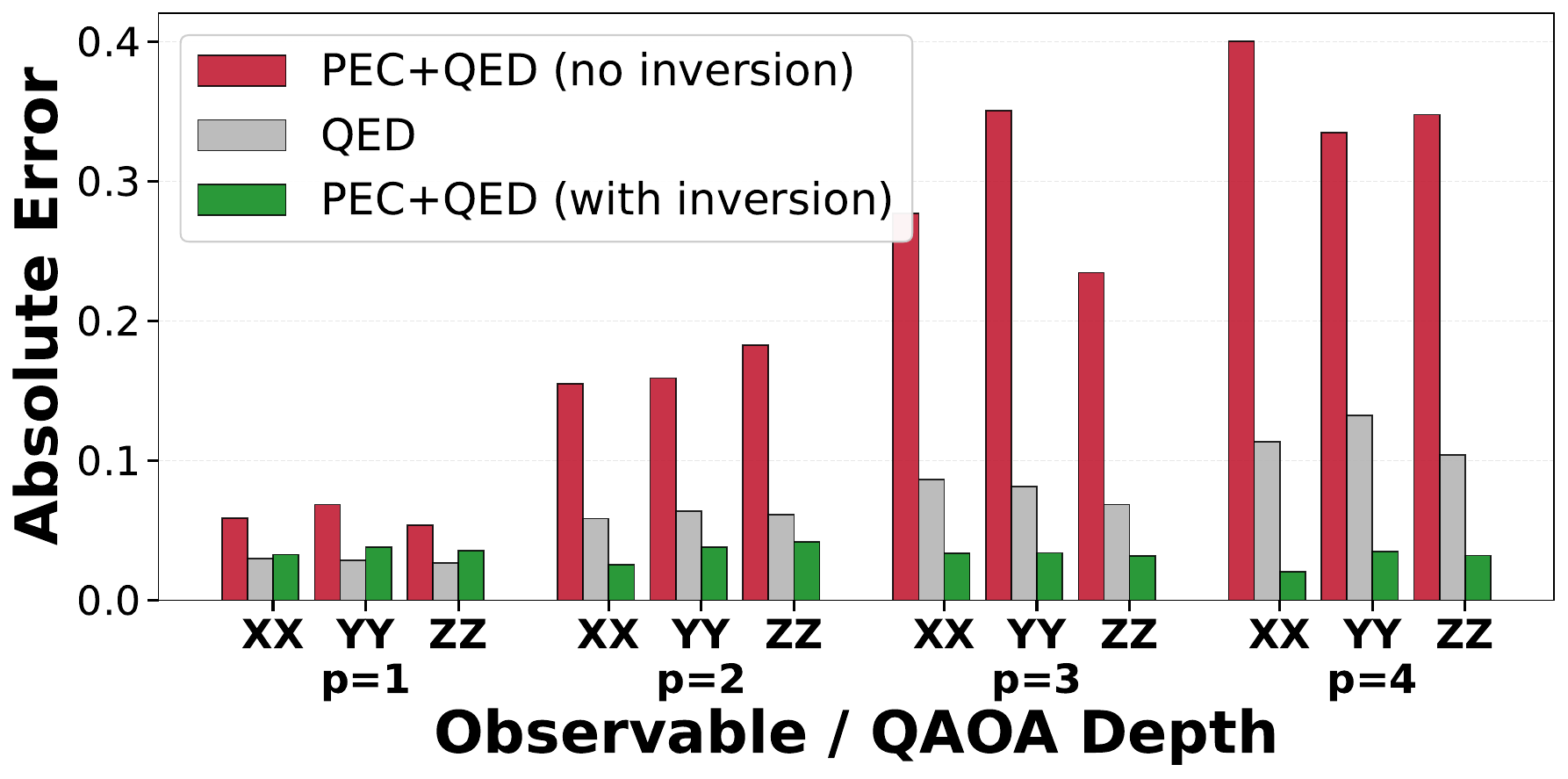}};
      \node[overlay, anchor=north west, xshift=-10pt, yshift=10pt, font=\bfseries]
        at (img.north west) {(f)};
    \end{tikzpicture}
  \end{minipage}
  \caption{\textbf{Hamiltonian-level validation of steady-state extraction} using QuTiP Lindblad simulation on the $[[4,2,2]]$ code ($p_2 = p_m = 10^{-3}$). \textbf{(a)}~Fidelity evolution across QED codes ($[[5,1,3]]$, $[[4,2,2]]$, $[[4,1,2]]$): all exhibit fast-decay transients in the first cycle followed by slow-decay dynamics, validating Section~\ref{sec:eigenstructure}. \textbf{(b)}~Logical-level characterization achieves $7.67\times$ lower prediction error than physical-level, demonstrating QED measurement error resilience. \textbf{(c)}~Steady-state extraction reduces per-cycle prediction error by $39.7\times$ versus naive tomography. \textbf{(d)--(f)}~End-to-end PEC+QED observable error under (d)~depolarizing, (e)~amplitude damping, and (f)~dephasing noise. Naive PEC degrades below QED-only; steady-state extraction enables up to $10.2\times$ bias reduction.}
  \label{fig:full_sim_results}
\end{figure*}

We validate steady-state extraction through Hamiltonian-level simulation of PEC on the $[[4,2,2]]$ QED code using the logical-level instantiation: both QPT characterization and superoperator inversion are performed in logical space. QED checks are inserted after every logical gate ($L=1$) across three noise models, constituting the most adversarial test of the protocol, as characterization errors compound over the maximum number of mitigated units at the smallest possible interval. Because this Hamiltonian-level circuit is serial at the logical level, each gate constitutes one layer.

\textbf{Simulation setup.} Using Hamiltonian-level QuTiP simulation, we characterize and mitigate logical gates for a two-qubit QAOA circuit implemented within a $[[4,2,2]]$ code patch. While full Lindblad master equation simulation limits us to this 5 qubit system (4 data + 1 ancilla), it provides fine-grained visibility into noisy gate and QED cycle behavior. In particular, Hamiltonian-level simulation enables us to test non-unital noise models (in this case, amplitude damping noise) and model errors that can occur within gates themselves, causing complex emergent dynamics even for simple error models.

\textbf{Circuit structure and implementation.}
Our QAOA circuits consist of logical $\overline{ZZ}(\gamma)$ and $\overline{R_x}(\beta)$ gates applied to two encoded qubits. Each gate reduces to a single physical two-qubit interaction~\cite{self2024protecting, jin2025iceberg}. We evaluate circuits across varying QAOA depths $p$, optimizing parameters for a two-vertex MaxCut instance. The optimized circuits produce ideal expectation values $\langle ZZ \rangle = -1$ and $\langle XX \rangle = \langle YY \rangle = 1$.

\textbf{PEC basis and characterization protocol.}
We employ the tensor-product extension of the one-qubit basis from~\cite{endo2018practical} to span the two-qubit CPTP space. For each logical gate, we compute the quasi-probability decomposition of the ideal gate in terms of its characterized noisy implementation and the basis operations. Each basis element is characterized via steady-state extraction (Section~\ref{sec:sse_protocol}).

\textbf{Characterization validation.} Before evaluating end-to-end PEC performance, we verify that steady-state extraction accurately characterizes the steady-state channel. Figure~\ref{subfig:inv_supremacy} compares the absolute fidelity prediction error for QED cycles in the slow-decay regime using naive single-cycle tomography versus steady-state extraction. Steady-state extraction reduces prediction error by $39.7\times$ per cycle, confirming that the protocol correctly isolates the steady-state dynamics that govern logical gate operation.

\textbf{End-to-end results.}
Figure~\ref{fig:full_sim_results}(d)--(f) presents the absolute error in observable expectation values $\langle XX \rangle$, $\langle YY \rangle$, and $\langle ZZ \rangle$ across QAOA depths for all three noise models. The two-qubit gate error and measurement error probabilities in each case were $10^{-3}$. We compare QED alone, QED with PEC using naive single-cycle tomography, and QED with PEC using steady-state extraction.

The results demonstrate two critical findings. First, PEC in conjunction with naive tomography, which inappropriately accounts for transient dynamics, produces \emph{worse} performance than QED alone. This highlights that attempting to combine PEC and QED without accounting for the unique transient error behavior of QED codes is actively harmful. Second, steady-state extraction successfully corrects for transient dynamics, enabling PEC to improve fidelity beyond QED alone across all three noise models and observables. In the $p=4$ QAOA circuits, the average bias reduction across observables was $4.21\times$ for the dephasing channel, $3.57\times$ for the depolarizing channel, and $6.37\times$ for the amplitude damping channel. In the best case, the mitigated circuit had bias $10.2\times$ lower than the QED-only baseline. The consistent advantage across all three noise channels confirms that our approach captures fundamental properties of QED channels rather than noise-specific artifacts.

These results establish that accounting for first-cycle position-dependent errors is critical for realizing the theoretical advantages of combining PEC with QED; without it, performance will be actively degraded.

\section{End-to-End PEC+QED Evaluation}\label{sec:e2e_eval}

\begin{figure}[t]
  \centering
  \includegraphics[width=\columnwidth]{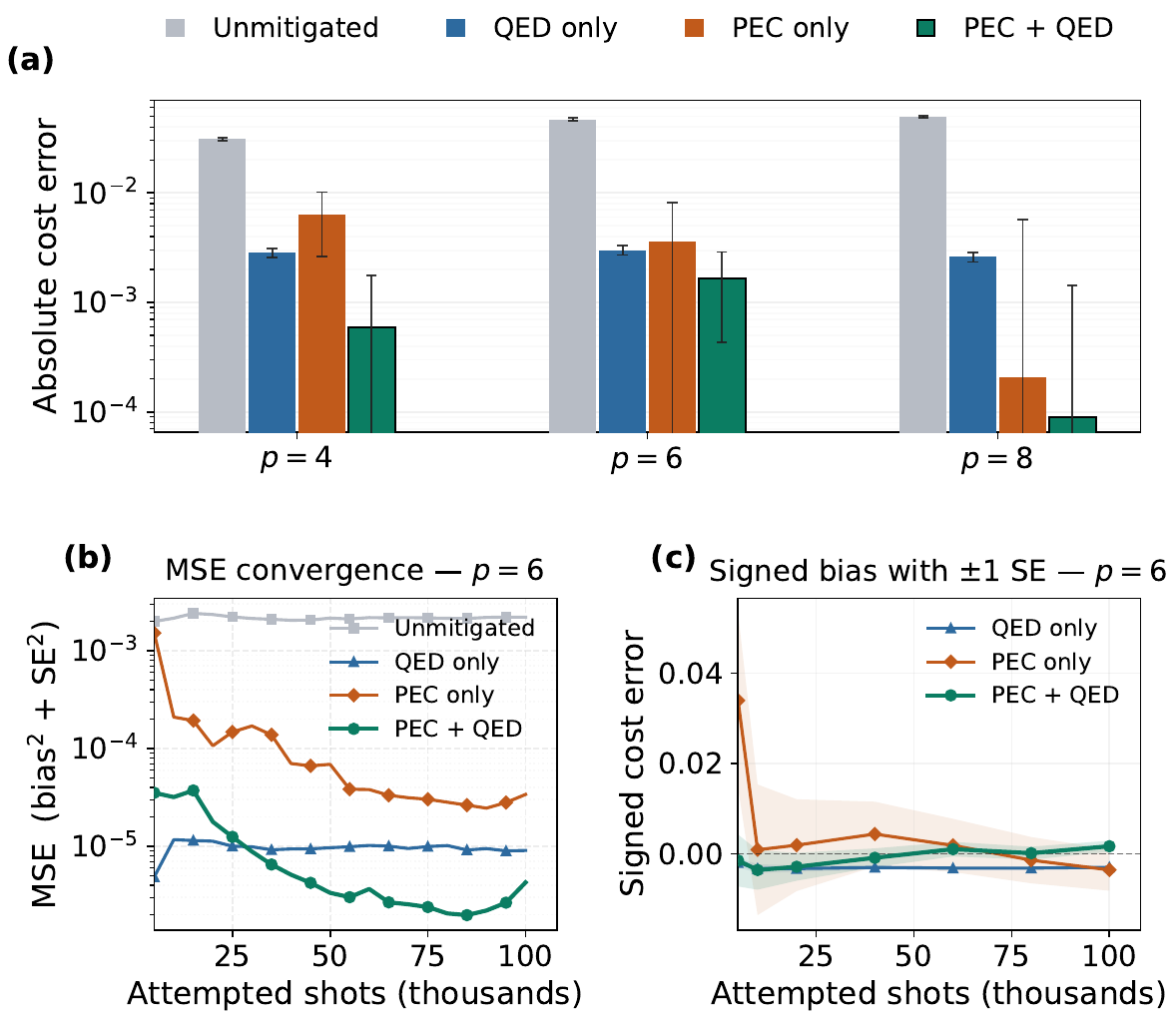}
  \caption{\textbf{End-to-end PEC+QED evaluation on the $[[6,4,2]]$ Iceberg code across QAOA depths $p{=}4$ to $p{=}8$.} \textbf{(a)}~Absolute cost error at $10^5$ attempted shots for all four protocols across three circuit depths. PEC+QED (green, black-outlined) achieves the lowest error at every depth. Error bars show $\pm 1$ standard error. \textbf{(b)}~MSE convergence for the $p{=}6$ configuration with midpoint and terminal QED checks. PEC+QED achieves the lowest MSE across the entire shot range. \textbf{(c)}~Signed cost error with $\pm 1$ standard error bands for $p{=}6$ (unmitigated baseline omitted to preserve scale). PEC+QED's estimate remains close to zero bias, while PEC-only exhibits larger variance. Numerical details in Table~\ref{tab:e2e_summary}.}
  \label{fig:e2e_results}
\end{figure}

We now evaluate the complete PEC+QED pipeline on the $[[6,4,2]]$ Iceberg code across QAOA depths $p{=}4$ to $p{=}8$.

\textbf{Setup.} We simulate a $[[6,4,2]]$ Iceberg code implementing an all-to-all QAOA MaxCut instance on 4 logical qubits at three circuit depths ($p{=}4, 6, 8$), using Cirq~\cite{cirq2024} for Monte Carlo simulation under depolarizing noise ($p_1 = 10^{-4}$, $p_2 = 10^{-3}$). All configurations use two QED checks: one at the midpoint of the circuit and one at the end. Fixing two checks while increasing circuit depth (controlled by $p$) implicitly grows the QED interval per mitigable unit, while also evaluating the efficacy of characterization and mitigation of both the transient and steady-state dynamics. With this QED schedule, the effective QED interval per mitigable unit is above the break-even threshold $L_{\text{thresh}}{=}4$ identified in Section~\ref{sec:sampling_adv}; for $p=4, 6, 8$ this interval is 20, 30, and 40 respectively. We compare four protocols: (1)~unmitigated, (2)~QED-only, (3)~PEC-only on physical qubits, and (4)~PEC+QED with steady-state extraction, each evaluated over a cumulative shot sweep up to $10^5$ attempted shots. 

\begin{table}[!t]
\centering
\setlength{\tabcolsep}{2.4pt}
\resizebox{\columnwidth}{!}{%
\begin{tabular}{@{}llccccc@{}}
\toprule
\textbf{Depth} & \textbf{Protocol} & \textbf{Abs.\ Err.} & \textbf{SE} & \textbf{MSE} & $\boldsymbol{\gamma}$ & \textbf{Acc.} \\
\midrule
\multirow{4}{*}{$p{=}4$}
 & Unmitigated  & $3.10{\times}10^{-2}$ & $8.7{\times}10^{-4}$ & $9.6{\times}10^{-4}$ & --- & --- \\
 & QED only     & $2.9{\times}10^{-3}$  & $2.8{\times}10^{-4}$ & $8.2{\times}10^{-6}$ & --- & 93.1\% \\
 & PEC only     & $6.4{\times}10^{-3}$  & $3.8{\times}10^{-3}$ & $5.5{\times}10^{-5}$ & 1.053 & --- \\
 & \textbf{PEC+QED} & $\mathbf{6.0{\times}10^{-4}}$ & $1.2{\times}10^{-3}$ & $\mathbf{1.7{\times}10^{-6}}$ & 1.005 & 93.9\% \\
\midrule
\multirow{4}{*}{$p{=}6$}
 & Unmitigated  & $4.69{\times}10^{-2}$ & $1.1{\times}10^{-3}$ & $2.2{\times}10^{-3}$ & --- & --- \\
 & QED only     & $3.0{\times}10^{-3}$  & $2.9{\times}10^{-4}$ & $9.1{\times}10^{-6}$ & --- & 92.0\% \\
 & PEC only     & $3.6{\times}10^{-3}$  & $4.6{\times}10^{-3}$ & $3.4{\times}10^{-5}$ & 1.080 & --- \\
 & \textbf{PEC+QED} & $\mathbf{1.7{\times}10^{-3}}$ & $1.2{\times}10^{-3}$ & $\mathbf{4.3{\times}10^{-6}}$ & 1.006 & 91.9\% \\
\midrule
\multirow{4}{*}{$p{=}8$}
 & Unmitigated  & $4.99{\times}10^{-2}$ & $1.1{\times}10^{-3}$ & $2.5{\times}10^{-3}$ & --- & --- \\
 & QED only     & $2.6{\times}10^{-3}$  & $2.7{\times}10^{-4}$ & $6.9{\times}10^{-6}$ & --- & 90.2\% \\
 & PEC only     & $2.1{\times}10^{-4}$  & $5.5{\times}10^{-3}$ & $3.0{\times}10^{-5}$ & 1.108 & --- \\
 & \textbf{PEC+QED} & $\mathbf{9.0{\times}10^{-5}}$ & $1.3{\times}10^{-3}$ & $\mathbf{1.8{\times}10^{-6}}$ & 1.006 & 90.2\% \\
\bottomrule
\end{tabular}%
}
\caption{\textbf{End-to-end evaluation summary at $10^5$ attempted shots.} MSE $= \text{bias}^2 + \text{SE}^2$. $\gamma$ is the PEC quasi-probability norm (mean absolute sample weight; --- indicates no PEC weighting). Acc.\ is the QED acceptance rate (fraction of shots passing both syndrome checks). PEC+QED achieves the lowest MSE at every depth. PEC-only's $\gamma$ grows from 1.053 ($p{=}4$) to 1.108 ($p{=}8$), while PEC+QED maintains $\gamma < 1.01$ at all depths.}
\label{tab:e2e_summary}
\end{table}

\textbf{Characterization.} We employ propagation-based error characterization in the style of Zhong et al.~\cite{zhong2025combining}, which reconstructs the error model by propagating known fault distributions through the circuit structure. Because this method directly produces physical-level superoperators, we apply steady-state extraction at the \textit{physical level}: the first-cycle transient is inverted out in the physical Hilbert space before projecting to logical space. As discussed in Section~\ref{sec:sse_protocol}, this physical-level approach is natural for the position-dependent error problem because the leakage states driving the transient are clearly resolved as distinct basis states, whereas at the logical level their effects are averaged into the logical channel.

\textbf{Evaluation metric.} We report two metrics. First, the absolute observable error at the full shot budget, since reducing expectation-value bias is the principal goal of PEC and QED in our context. Second, mean squared error $\text{MSE} = \text{bias}^2 + \text{SE}^2$, where bias is the signed deviation of the estimate from the ideal observable value and SE is the standard error. MSE jointly penalizes systematic error and statistical uncertainty, enabling fair comparison between protocols that trade bias for variance: reporting bias alone would favor high-variance estimators, while reporting variance alone would ignore the residual errors that motivate mitigation.

\textbf{Results.} Figure~\ref{fig:e2e_results}(a) and Table~\ref{tab:e2e_summary} present the endpoint comparison at $10^5$ attempted shots across all three depths. PEC+QED achieves the lowest absolute observable error and MSE at every depth. As the circuit depth, governed by $p$, and thereby the interval grow, the total overhead $\gamma$ for the PEC-only configuration increases from 1.053 at $p{=}4$ to 1.108 at $p{=}8$, reflecting the growing noise per mitigable unit that physical-qubit PEC must cancel. The $\gamma$ overhead for PEC+QED, by contrast, grows much more slowly, increasing from 1.005 to 1.006 from $p{=}4$ to $p{=}8$, because QED limits the per-interval logical error rate. The widening ratio of PEC-only to PEC+QED $\gamma$ with depth is consistent with the sampling advantage condition (Equation~\ref{eq:efficiency_condition}): each larger interval pushes further into the favorable regime, producing a growing gap in sampling efficiency.

Figure~\ref{fig:e2e_results}(b) shows MSE convergence at $p{=}6$, where midpoint and terminal QED checks produce two QED cycles. PEC+QED achieves the lowest MSE across the entire shot range, converging to $4.3 \times 10^{-6}$ at $10^5$ shots, compared to $3.4 \times 10^{-5}$ for PEC-only ($7.9\times$ higher) and $9.1 \times 10^{-6}$ for QED-only ($2.1\times$ higher). Panel~(c) shows the signed error traces at $p{=}6$: PEC-only exhibits larger variance than PEC+QED; QED-only retains a residual bias of $-0.003$ from undetectable errors; PEC+QED centers near zero bias while maintaining lower variance than PEC-only. The PEC+QED acceptance rates across depths range from 93.9\% ($p{=}4$) to 90.2\% ($p{=}8$), confirming that the postselection overhead remains moderate even at the deepest circuits.

These results validate the complete co-designed pipeline. Physical-level steady-state extraction successfully isolates the first-cycle transient across both QED cycles, enabling PEC to reduce bias below QED-only baselines rather than degrading it. The scaling trend across intervals confirms the sampling advantage prediction of Section~\ref{sec:sampling_adv}. The central practical result is that, for a fixed shot budget, the co-design of interval optimization and steady-state extraction achieves far lower observable error than either PEC or QED alone.

\section{Resource Model}\label{sec:resource_model}

Deploying PEC+QED trades modest overhead across four resource dimensions for exponential sampling savings derived in Section~\ref{sec:sampling_adv}. Space and gate time are structural costs, while characterization maps to the offline phase and shot budget to the online phase of the execution architecture (Figure~\ref{fig:intro-exec}).

\textbf{Space overhead.} High-rate Iceberg codes incur minimal physical qubit overhead: $1.5\times$ for the $[[6,4,2]]$ code and $1.2\times$ for the $[[12,10,2]]$ code, compared to $9\times$ for the $[[9,1,3]]$ surface code. This favorable encoding rate is both a prerequisite for sampling advantage and a practical advantage for near-term deployments where physical qubits are scarce.

\textbf{Gate execution time.} On Iceberg codes, single-qubit logical rotations are implemented using physical two-qubit interactions (e.g., a logical $R_x$ rotation is implemented through an $XX$ interaction~\cite{self2024protecting, jin2025iceberg}), increasing layer latency when such rotations are executed serially. However, gauge fixing~\cite{niroula2026digital} mitigates this by introducing proxy qubits that allow single-qubit logical rotations to be parallelized, reducing the single-qubit logical layer depth to a low constant. Because gauge-fixed compilation is not evaluated in this work, our advantage claims are scoped to sampling efficiency (Section~\ref{sec:sampling_adv}); quantifying end-to-end runtime improvement remains future work.

\textbf{Characterization cost.} Steady-state extraction requires one-time offline characterization of the initialization superoperator and each gate type. The method determines scaling: direct logical-level tomography incurs exponential cost in logical qubits per patch, while propagation-based methods~\cite{zhong2025combining} leverage circuit structure and physical-level error models to avoid this. In either case, the noise model is a hardware property; characterization is a fixed upfront investment that amortizes across all circuits on a given device.

\textbf{Online shot budget.} During execution, the total shot count is governed by the quasi-probability norm $\gamma$ and the QED acceptance rate. The sampling advantage condition (Equation~\ref{eq:efficiency_condition}) ensures this cost favors PEC+QED over physical-qubit PEC for appropriately chosen codes and intervals.

\section{Related Work}
\label{sec:related_work}

Zhong et al. \cite{zhong2025combining} also combine QED with PEC on Iceberg codes. Their propagation-based characterization approach, which we adopt with physical-level steady-state extraction in our end-to-end evaluation (Section~\ref{sec:e2e_eval}), effectively constructs error models. Two key differences distinguish our work. First, Zhong et al. apply QED only once at the end of the circuit rather than at multiple points throughout, which is why they do not encounter the position-dependent first-cycle transient that arises with repeated QED cycles. Their operating point is consistent with our interval optimization results, corresponding to a regime where the efficiency condition (Equation~\ref{eq:efficiency_condition}) is naturally satisfied. Second, while they demonstrate PEC+QED in practice, they do not quantify \textit{when, why, or under what conditions} sampling advantage is expected; we define architectural parameters for reliably doing so. We extend their propagation-based approach with steady-state extraction to handle the multi-QED-cycle case.

Suzuki et al. \cite{suzuki2022quantum} proposed integrating PEC and surface code QEC, showing PEC can effectively increase code distance and T-gate count by mitigating decoding and gate decomposition errors. Their framework assumes logical noise is Markovian (position-independent) and stochastic Pauli. Our work differs in two respects. First, we operate in the nearer-term quantum error \emph{detection} (QED) regime rather than the QEC regime, and answer the \textit{QED-specific} question of when PEC applied to QED provides sampling advantage. Our characterization of each mitigable unit (Section~\ref{sec:sse}) captures both error categories in \cite{suzuki2022quantum}, and further captures all logical-level errors. Second, we show that the position-independence assumption does \emph{not} hold in the QED setting, and introduce steady-state extraction to address this (Section~\ref{sec:sse}).

Kwiatkowski et al. \cite{kwiatkowski2025constructing} show stabilizer-code dynamics can be approximated by position-independent channels with error $O(G\epsilon^K)$ for $K$ syndrome extractions. This explicitly allows for (but does not itself predict) position-dependent behavior for small $K$, complementing our findings.

Independent work by Ziyad et al.~\cite{ziyad2025emergent} formalizes logical non-Markovianity via a ``button-theoretic'' definition of gate composability and proves any $d \geq 3$ stabilizer code with noisy syndrome readout violates it. We identify the specific dynamical mechanism driving this behavior: asymmetric leakage injection versus removal in the first QED cycle (Section~\ref{sec:physical_interp}), yielding a clean spectral separation into one $\Theta(p)$ fast mode and $1 - \Theta(p)$ slow modes (Section~\ref{sec:eigenstructure}) that that localizes the dominant transient effect to the first detection cycle. This structural understanding enables our steady-state extraction protocol (Section~\ref{sec:sse}), a concrete superoperator-inversion procedure that isolates the steady-state channel and, to our knowledge, is the first method to make PEC compatible with repeated QED cycles.

\section{Discussion}\label{sec:discussion}

This work occupies a specific and timely position in quantum computing, bridging the gap between the NISQ era, where algorithms run directly on physical qubits, and the eventual fault-tolerant quantum computing (FTQC) era, where scalable QEC supports arbitrarily long computations.

QED is already demonstrating practical value on real hardware \cite{he2025performance, reichardt2024fault, rines2025demonstration, linke2017fault, vuillot2017error}. Layering PEC on top of these QED implementations requires no additional hardware engineering: it only needs accurate channel characterization and classical post-processing.

By contrast, full QEC demands substantially more engineering infrastructure~\cite{Dennis2002,Chen2021,google2023suppressing}: sub-microsecond feedback, magic state distillation, and large qubit overhead. Whereas a distance-3 surface code requires 9 physical qubits per logical qubit (a $9\times$ overhead)~\cite{Dennis2002} and real-time decoding, the $[[12,10,2]]$ Iceberg code only requires a $1.2\times$ overhead with no feedback loop, while PEC+QED reduces the sampling overhead of using such codes for mitigation.

The principles in this work extend to QEC, where distinct decoder branches can be characterized and mitigated using our techniques. The efficiency tradeoffs derived in Section~\ref{sec:sampling_adv} are relevant to QEC settings that include post-selection, such as soft-information decoding that discards low-confidence shots~\cite{meister2024soft,smith2024mitigating,dinca2025decoderconfidence} or erasure-dominated architectures that enable higher-distance QEC~\cite{teoh2023dualrail, wu2022erasureconversion}. Accounting for logical position-dependent errors via steady-state extraction will remain necessary when conditioning on results from syndrome extraction.

This points toward a progressive roadmap: from pure PEC on physical qubits in the NISQ era, to PEC+QED as demonstrated in this work, to PEC+QED+QEC as hardware matures, and ultimately to PEC+QEC in the fault-tolerant era. Each step along this trajectory incrementally reduces the mitigation burden as the hardware's native error handling improves, and the principles developed here inform design at every stage, from the sampling advantage condition and interval optimization to steady-state extraction.

\section{Conclusion}
\label{sec:conclusion}

We have shown when and how to combine Probabilistic Error Cancellation with Quantum Error Detection: a sampling advantage condition identifies high-rate Iceberg codes with optimized QED intervals as the favorable regime, while steady-state extraction isolates a position-dependent first-cycle transient that otherwise causes PEC to degrade below QED-only baselines, reducing observable bias by up to $10.2\times$. End-to-end evaluation on a $[[6,4,2]]$ QAOA instance at depths $p{=}4$ through $p{=}8$ confirms both contributions jointly: PEC+QED achieves $2$--$11\times$ lower absolute error than PEC alone and up to $31\times$ lower MSE, with the advantage compounding as interval depth grows. These results demonstrate that error detection and error mitigation are complementary techniques whose architectural co-design yields practical and growing advantages for near-term quantum computation.

\begin{acknowledgments}
This work was supported by the Defense Advanced Research Projects Agency (DARPA) Imaging Practical Applications for a Quantum Tomorrow (IMPAQT) program under Agreement No. HR0011-23-3-00019. This material is also supported by R.S.K.'s Graduate Research Fellowship from the National Science Foundation. R.S.K. thanks Roberto C. Cohen for feedback on the figures and visual assets.

External interest disclosure: RJS and YD are consultants and equity holders for D-Wave Quantum, Inc.
\end{acknowledgments}

\appendix

\section{Transition Matrix Construction and Spectral Analysis}\label{app:tm}

This appendix provides the formal transition matrix framework summarized in Section~\ref{sec:eigenstructure}. We construct the transition matrix, derive its eigenstructure via perturbation theory, and detail the basis-state reduction used for the $[[4,2,2]]$ simulations.

\subsection{Full Transition Matrix Construction}\label{app:tm_construct}

To trace the origin of the position-dependent behavior, we need a framework that tracks how quantum state populations evolve through successive QED cycles. We model each noisy QED cycle as a stochastic process on Hilbert-space basis states, constructing a transition matrix whose spectral properties will reveal the two-timescale dynamics. Our basis comprises $2^k$ logical codewords plus $2^n - 2^k$ leakage states corresponding to detectable error syndromes, augmented with $n-k$ ancilla qubits used for syndrome measurement. In the \textit{noiseless} limit, syndrome extraction acts as the identity on logical states while flipping ancilla registers to signal detected errors in leakage states. This defines a transition matrix $T_0$ with block structure separating the logical subspace (which is preserved) from leakage states (which are mapped to flagged ancilla configurations).

\begin{figure}[!t]
  \centering

  \begin{subfigure}{\columnwidth}
    \centering
    \begin{tikzpicture}[baseline]
      \node[anchor=north west,inner sep=0] (img)
        {\includegraphics[width=0.95\columnwidth]{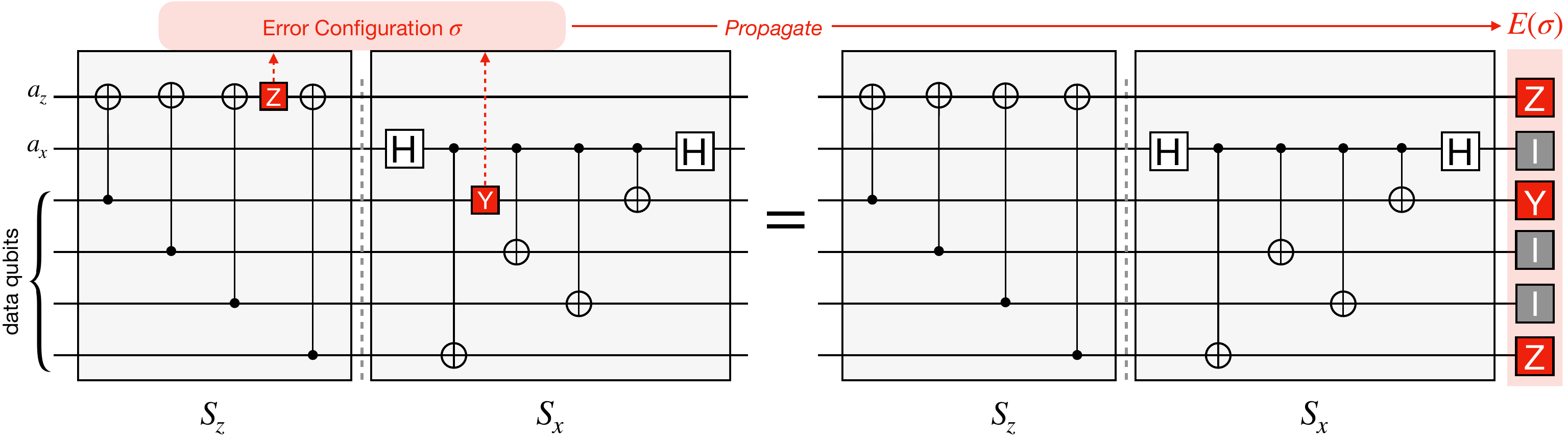}};
      \node[overlay, anchor=north west, xshift=-10pt, yshift=12pt, font=\bfseries]
        at (img.north west) {(a)};
    \end{tikzpicture}
  \end{subfigure}

  \begin{subfigure}{\columnwidth}
    \centering
    \begin{tikzpicture}[baseline]
      \node[anchor=north west,inner sep=0] (img)
        {\includegraphics[width=0.95\columnwidth]{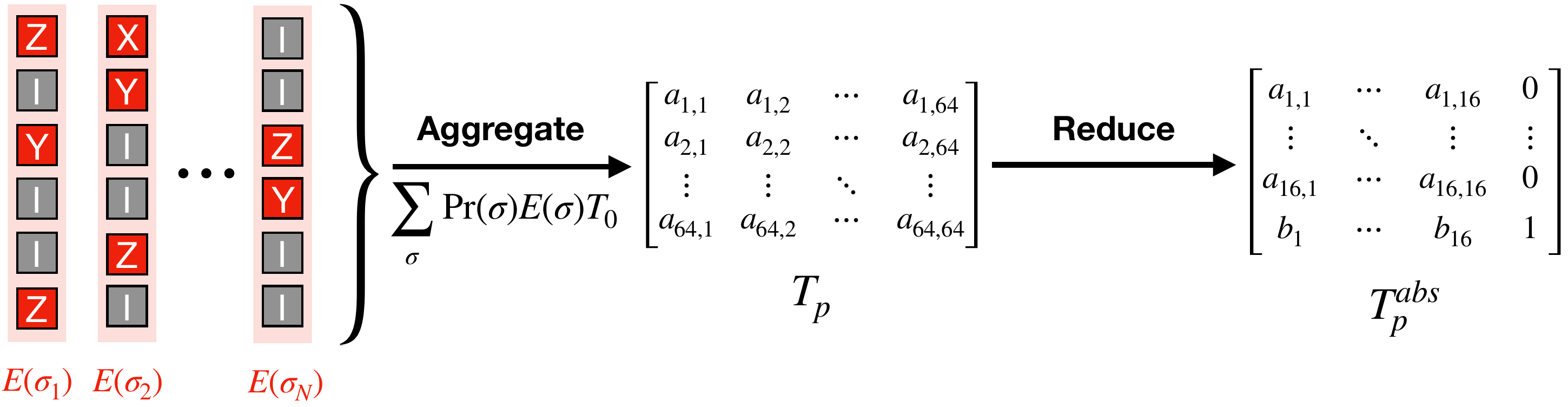}};
      \node[overlay, anchor=north west, xshift=-12pt, yshift=12pt, font=\bfseries]
        at (img.north west) {(b)};
    \end{tikzpicture}
  \end{subfigure}

  \caption{\textbf{(a)} Computation of $E(\sigma)$ for a particular error configuration $\sigma$ for the $[[4,2,2]]$ code. The $Z$ and $Y$ errors in their respective locations comprise one example of $\sigma$.
  \textbf{(b)} Taking a convex combination over errors $E(\sigma)$ with associated probabilities yields $T_p$, which is further reduced to $T_p^{\text{abs}}$ for Markov chain modeling.}
  \label{fig:tm_construct}
\end{figure}

To introduce noise, we insert single-qubit Pauli errors at some subset of the $K$ error locations throughout the syndrome extraction circuit, each occurring independently with probability $p$. Since the extraction circuit is Clifford, any Pauli fault pattern $\sigma$ propagates to produce an effective Pauli operator $E(\sigma)$ acting at the end of the circuit, as shown in Figure~\ref{fig:tm_construct}(a). An error configuration $\sigma$ with weight $|\sigma|$ occurs with probability $(1-p)^{K-|\sigma|}(p/3)^{|\sigma|}$. Averaging over all possible fault configurations, the noisy transition matrix becomes $T_p = \sum_{\sigma \in \{I,X,Y,Z\}^\Lambda} \Pr(\sigma) \, E(\sigma) \, T_0$. This construction is visualized in Figure~\ref{fig:tm_construct}(b).

\textbf{Postselection via absorbing reduction.}
Experimental postselection discards any trajectory that flags a syndrome. We model this by aggregating all flagged ancilla states into a single absorbing reject state, yielding the reduced matrix $T_p^{\mathrm{abs}}$ acting on accepted states plus the absorbing state.

\textbf{Multi-round dynamics and ASC.}
Starting from initial state $v^{(0)} = e_0$ (the first logical basis state), after $r$ QED cycles we have $v^{(r)} = (T_p^{\mathrm{abs}})^r v^{(0)}$. The last component $v^{(r)}_{-1}$ tracks cumulative rejection probability; survival probability is $S(r) = 1 - v^{(r)}_{-1}$. The \textbf{Accepted State Correctness (ASC)} is then $\mathrm{ASC}(r) := \frac{v^{(r)}_0}{S(r)} = \frac{v^{(r)}_0}{1 - v^{(r)}_{-1}}.$ This represents the conditional probability of remaining in the correct logical state given survival.

\subsection{Eigenstructure Derivation}\label{app:eigenstructure}

In the noiseless limit ($p \to 0$), the absorbing reduction of $T_0$ has spectrum $\sigma(T_0^{\mathrm{abs}}) = \{1, \ldots, 1\}_{2^k + 1} \cup \{0, \ldots, 0\}_{2^n - 2^k}$, where the eigenvalues of 1 correspond to the $2^k$ logical states plus the absorbing reject state, and the zero eigenvalues correspond to leakage states eliminated in one cycle.

For small but nonzero error probability $p$, standard perturbation theory~\cite{bauer1960norms,seneta1981nonnegative} shows that $O(p)$ noise shifts eigenvalues by $O(p)$ from their noiseless values. Combined with the stochastic property and absorbing structure, this yields:
\begin{equation}
    \sigma(T) = \{1\} \cup \{1 - \alpha_i p + O(p^2)\}_{i=1}^{2^k} \cup \{\beta_j p + O(p^2)\}_{j=1}^{2^n - 2^k},
\end{equation}
where $\alpha_i, \beta_j > 0$ are code- and circuit-specific constants. This spectrum defines three groups: one exactly stationary mode (the absorbing state), $2^k$ ``slow'' modes with eigenvalues $1 - O(p)$, and $2^n - 2^k$ ``fast'' modes with eigenvalues $O(p)$.

Because $T$ is diagonalizable, the state after $r$ rounds evolves as a linear combination of eigenmodes, each weighted by $\lambda_i^r$. Fast modes with eigenvalues $O(p)$ contribute significantly in the first QED cycle but vanish within $O(1)$ rounds. Subsequent cycles are dominated by slow modes with eigenvalues $1 - O(p)$, producing the steady-state decay behavior validated in Section~\ref{sec:tm_sim}.

\subsection{Basis-State Reduction for the $[[4,2,2]]$ Code}\label{app:422_reduction}

For the $[[4,2,2]]$ code, we perform an additional reduction on the basis states of the Hilbert space, classifying states into four classes: detected error states ($C_0$), the correct logical state ($C_1$), incorrect logical states ($C_2$), and undetected leakage states ($C_3$). This is possible due to the symmetric nature of the $[[4,2,2]]$ code; any leakage state is equidistant from all logical states, and vice versa. We label the resultant matrix $T_p^C$ and its eigenvalues $\lambda_i$ where $\lambda_i$ corresponds to the class $C_i$.

\section{Measurement Error Resilience at Characterization Time}\label{app:meas_error}

\begin{figure}[!t]
    \centering
    \includegraphics[width=\columnwidth]{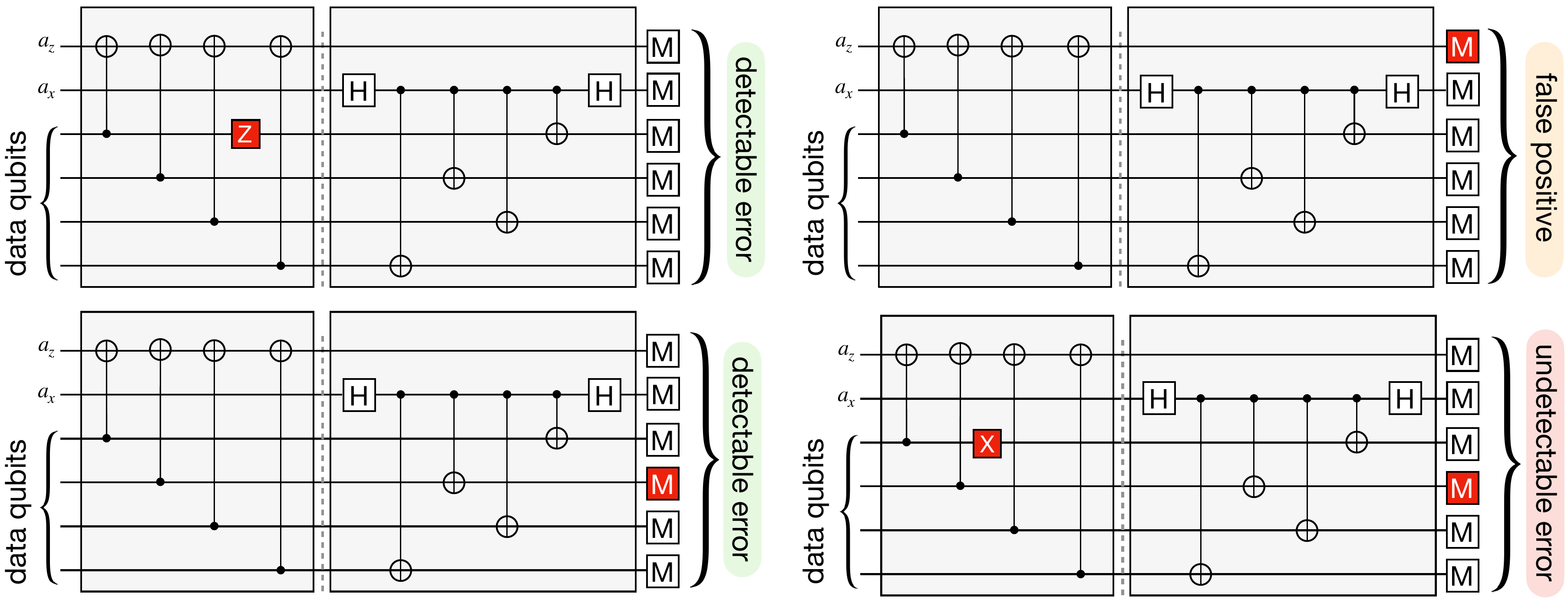}
    \caption{\textbf{Measurement error detection during QED characterization.} QED codes suppress measurement errors by converting them into false positives or detectable rejections. Isolated data qubit errors and data qubit measurement errors (top and bottom left) are detectable as they produce shots outside the code space. Isolated ancilla measurement errors (top right) result in false positives that do not degrade post-selected fidelity. Undetectable errors (bottom right) require \textit{coincident} syndrome extraction and measurement errors, making measurement error contributions negligible compared to physical-level characterization.}
    \label{fig:meas_err_detection}
\end{figure}

QED codes provide a structural property that benefits any characterization protocol: measurement errors are suppressed within accepted shots. In a single-round QED protocol, each logical gate is followed by syndrome extraction, and a shot is accepted only if all stabilizer checks return trivial outcomes. Suppose each stabilizer readout has an independent error rate $p_m$. A single flipped outcome on the data qubits causes a \emph{false positive} rather than corrupting the logical data that survive. Thus, first-order measurement faults appear primarily as reduced acceptance rather than bias within accepted shots. The dominant residual effects are higher order: (i) \emph{masked faults}, where a data error is concealed by a simultaneous readout flip, with probability $O(\epsilon_p p_m)$; and (ii) \emph{logical readout flips}, which require errors with probability $O(p_m^{\,d})$ ~\cite{Terhal2015}. Across leading hardware modalities (superconducting, trapped-ion, and neutral-atom), physical measurement error rates satisfy $p_m \lesssim 5\epsilon_p \leq 10^{-2}$~\cite{google2025quantum,ransford2025helios,bluvstein2024logical}, so the masked-fault contribution $O(\epsilon_p\, p_m)$ is suppressed by a factor of $\sim\!\epsilon_p$ relative to the unencoded measurement error and the logical readout term $p_m^d$ remains $30$--$1000\times$ below $\epsilon_p$ even for $d{=}2$. For very large measurement error rates $p_m \gg \sqrt{\epsilon_p}$, the effect of measurement error can become comparable to the two-qubit error rate on accepted shots, but this regime is unobserved in today's frontier experiments. Therefore, in practice the measurement error's effect on characterization remains definitively second order relative to the dominant source of noise. The mechanism for measurement error suppression in characterization is visualized with specific cases in Figure~\ref{fig:meas_err_detection}.

This measurement error suppression is what enables the superoperator inversion in steady-state extraction (Section~\ref{sec:sse_protocol}) to cleanly isolate the fast-decay transient. This property is most consequential for the logical-level approach, where QPT-based inversion requires that $S_{\text{init}}$ accurately reflects logical-state dynamics rather than measurement artifacts. For the physical-level approach using propagation-based methods, SPAM errors can additionally be handled by efficient physical-level techniques such as cycle benchmarking, which operate on individual physical qubits rather than the full logical space. Physical-level characterization thus benefits from both the intrinsic measurement error suppression of QED codes and the availability of efficient physical-level SPAM-handling techniques. This combination makes the physical-level approach particularly practical for high-rate codes.

To quantify this advantage, we compare logical-level characterization with steady-state extraction on the $[[4,2,2]]$ code against physical-level characterization on an equivalent system with identical noise and measurement error rates using Hamiltonian-level QuTiP simulation. As shown in Figure~\ref{subfig:qed_meas_supremacy}, logical-level characterization achieves $7.67\times$ lower fidelity prediction error per cycle, confirming that the structural measurement error suppression of QED codes translates into a concrete characterization advantage.

\bibliographystyle{apsrev4-2}
\bibliography{refs}

\end{document}